\documentclass[journal,twocolumn,final]{IEEEtran}
\usepackage[utf8]{inputenc}
\usepackage[T1]{fontenc}
\usepackage{amssymb}
\usepackage{amsmath}
\usepackage{mathtools}
\usepackage{algorithm}
\usepackage{algpseudocode}
\usepackage{graphics}
\usepackage{epsfig}
\usepackage{cite}
\usepackage{subfigure}
\usepackage{soul,xcolor}
\usepackage{amsfonts}
\usepackage{graphicx}
\usepackage{multirow}
\usepackage{textcomp}	
\usepackage{color}
\IEEEoverridecommandlockouts

\begin{document}

\title{Deep Reinforcement Learning Aided Monte Carlo Tree Search for MIMO Detection}

\author{Tz-Wei Mo, Ronald Y. Chang, \IEEEmembership{Member,~IEEE}, and Te-Yi Kan
\thanks{This work was supported in part by the Ministry of Science and Technology, Taiwan, under Grants MOST 106-2628-E-001-001-MY3 and MOST 109-2221-E-001-013-MY3.}
\thanks{The authors are with the Research Center for Information Technology Innovation, Academia Sinica, Taipei, Taiwan (e-mail: tzwmo1998@gapp.nthu.edu.tw, rchang@citi.sinica.edu.tw, dexter.ty.kan@gmail.com).}}

\maketitle

\begin{abstract}
This paper proposes a novel multiple-input multiple-output (MIMO) symbol detector that incorporates a deep reinforcement learning (DRL) agent into the Monte Carlo tree search (MCTS) detection algorithm. We first describe how the MCTS algorithm, used in many decision-making problems, is applied to the MIMO detection problem. Then, we introduce a self-designed deep reinforcement learning agent, consisting of a policy value network and a state value network, which is trained to detect MIMO symbols. The outputs of the trained networks are adopted into a modified MCTS detection algorithm to provide useful node statistics and facilitate enhanced tree search process. The resulted scheme, termed the DRL-MCTS detector, demonstrates significant improvements over the original MCTS detection algorithm and exhibits favorable performance compared to other existing linear and DNN-based detection methods under varying channel conditions.
\end{abstract}


\IEEEpeerreviewmaketitle

\section{Introduction}

It is evident that the usage of mobile communication continues to rise throughout the years. To match the increasing needs, newer generations of communication systems are being developed to reach higher data rate and more reliable transmissions. Currently, as the world shifts to fifth-generation (5G) communication systems, a key technology lies in multiple-input multiple-output (MIMO) systems, where multiple transmitting and receiving antennas are used simultaneously to pass data from the transmitter to the receiver. MIMO systems are essential in 5G networks since they employ multiple antennas to overcome multipath fading problems and increase spectral efficiency through achieving space diversity. One problem that arises in such a system, however, is the recovery of symbols at the receiver, namely, symbol detection, as the computational complexity grows exponentially with the number of transmit antennas and becomes NP-hard \cite{verdu1989} for an optimal solution. Many previous works have tried to reduce the complexity by using suboptimal receivers, which improves efficiency at the cost of various degrees of accuracy degradation depending on the method. In this work, we propose to solve the MIMO detection problem using a machine learning, specifically deep reinforcement learning (DRL), assisted tree search, as recent advances in neural network driven methods have shown success in numerous fields of engineering.   

\subsection{Related Work}

It has long been known that the maximum likelihood (ML) detector is the optimum detector for the MIMO detection problem \cite{shaoshi2015mimo}. However, its high computational complexity prevents it from being practical in setups with large numbers of antennas and/or high modulation orders. Suboptimum detectors have been proposed since the 1960s \cite{shaoshi2015mimo}, where initially linear methods such as the zero forcing (ZF) detector \cite{cheng2007zf} and minimum mean square error (MMSE) detector \cite{ChangChung2012} were developed. Although linear MIMO detectors in general have poorer performance, they have been shown to have ML-approaching performance as the number of antennas in the system increases, due to a ``channel-hardening" effect, a result of the Mar\v{c}enko and Pastur law \cite{marcenko1967channelharden}. However, at these massive MIMO setups, the complexity for high-order matrix inversion in linear methods also becomes prohibitive, and as such, matrix inversion approximation methods were proposed \cite{Michael2013, dengkui2015, byunggi2015} to accelerate the inversion process. 

Another set of detectors make use of nonlinear methods to tackle the MIMO detection problem. Interference cancellation methods \cite{andrew2005}, such as successive interference cancellation (SIC) \cite{Bittner2006,Jinho2009}, try to eliminate the effects of intersymbol interference (ISI) by using iterative algorithms to recover one symbol at a time while treating other symbols as noise. Local search methods, such as likelihood ascent search (LAS) \cite{Peng2010} or reactive tabu search (RTS) \cite{Tanumay2010}, start with an initial solution, which can be the result from linear detectors, then searches in a defined neighborhood for a better solution, where a local optimum is selected after certain iterations. Local search methods, while having the benefit of a tunable complexity with detection performance-speed tradeoff, are prone to the local minima problem if not given an extensive search. Tree search algorithms are also used in detectors \cite{Jong2005, Atsushi2008, Ronald2012, Ronald2012_2, Ronald2012_3}, where the MIMO detection problem is formulated as a decision tree and a symbol is recovered at each layer of the tree. In \cite{jienan2017}, a statistical approach with the Monte Carlo tree search (MCTS) algorithm was proposed with hardware acceleration to recover the transmitted symbols at large MIMO setups. Other nonlinear methods such as belief propagation (BP) \cite{5503188,Junmei2015}, semidefinite relaxation (SDR)\cite{Jalden2008}, and approximate message passing (AMP) \cite{Jeon2015} are all able to achieve good performance under many practical scenarios, while having lower complexity than the ML detector. Nonlinear methods usually have better performances than linear methods by using iterative and recursive algorithms. However, most nonlinear methods mentioned above require some parameter tuning, which can be difficult to determine.

In recent years, as hardware advancements increase computation capability along with the development of powerful optimization algorithms, deep neural networks (DNN) have become feasible and are finding applications in many research fields. Work in \cite{Chen2019} modeled MIMO detection as a classification problem which they proposed to solve using two machine learning methods: a DNN approach and a convolutional neural network (CNN) approach. Another way neural networks can be applied in MIMO detection is to use DNNs in place of traditional iterative algorithms where each iteration can be viewed as a layer of artificial neurons and the unfolded algorithm becomes a DNN. DetNet \cite{Samuel2019} uses DNN to unfold a projected gradient descent algorithm to try to reach the optimum solution. A related work \cite{Vincent2018} enhances DetNet by using a self-designed multilevel sigmoid function for better symbol classification and introduces a twin-network architecture to overcome an initialization problem with a random forest tree approach. WeSNet proposed in \cite{mohammad2020} improves upon the work in \cite{Samuel2019} by introducing a weight scaling framework which can further lower the network size and allows the network to self-adjust to the detection complexity. In \cite{jin2020}, a parallel detection network (PDN) is developed by using parallel DNNs without connection to solve a problem in \cite{Samuel2019} where increasing the number of detection layers does not significantly improve performance. Aside from unfolding the projected descent algorithm, in \cite{Nir2020}, DNN was incorporated into an existing soft SIC algorithm, and was shown to reach near-ML performance without channel state information (CSI) in certain cases. In \cite{Jianyong2020}, a long short term memory (LSTM) recurrent neural network (RNN) was proposed, where the RNN was shown to learn its own decision algorithm for detection and was exhibited to achieve near-ML performance in both fixed and varying channel cases for quadrature phase shift keying (QPSK) modulation. In \cite{Tan2020}, DNNs were used to unfold two BP algorithms with two separate networks, namely, DNN-BP and DNN-MS; and in \cite{Hengtao2018} layers of DNN are used to replace iterations in an orthogonal AMP (OAMP) algorithm; both works have shown DNN versions to outperform their traditional counterparts.  

The DNN approaches mentioned above all show promising results and improvements over traditional methods. The network parameters are also tuned implicitly during the training process via backpropagation without the need for manual adjusting. However, since these neural network methods are trained using supervised learning, they face a {\it labeling dilemma} over whether to use the ML solutions or the actual transmitted symbol vectors as labels during training \cite{Vincent2018}. It was advocated in \cite{Vincent2018} that ``the label that should be used for a given (received signal) $y$ is what would have been decoded by the optimal decoder, not the transmitted sequence'' for the reason that in order for the DNN to be tuned to become a quasi-ML detector, it needs to learn the correct decision boundaries of an optimal decoder. Using the actual transmitted sequence as labels may cause the network to learn undesirable boundaries and therefore result in performance degradation. The ML solution required for training, however, is almost impossible to obtain for large-scale MIMO systems. Thus, it was suggested \cite{Vincent2018} that the neural networks be trained under a fixed SNR corresponding to a $10^{-2}$ error probability in the actual transmitted sequence (i.e., $1$ out of $100$ symbols is mislabeled), such that the neural networks can still learn sufficient decision boundary information using the actual transmitted sequence instead of the desired ML labeling. This suggestion however only has empirical basis, and, to our knowledge, there still lacks definite answers to this dilemma.

\subsection{Main Contributions}

Inspired by DNNs and various machine learning architectures, in this work, we propose to use deep reinforcement learning (DRL) to work alongside the MCTS algorithm for MIMO detection. Unlike the aforementioned DNN approaches, we do not unfold existing algorithms; instead, our DNNs are used to improve the MCTS detection algorithm by providing useful statistics given the channel information and the received symbol vector. Our proposed method can be trained to detect under various antenna/modulation settings with favorable performance. Our main contributions are summarized as follows:
\begin{enumerate}
\item We propose to incorporate a DRL architecture into the traditional MCTS MIMO detection algorithm. The MIMO detection problem is formulated into a Markov decision process (MDP), which is then optimized through a self-play process where the DRL agent learns its own detection algorithm by generating its own training data to correctly recover the transmitted symbols.
\item The proposed scheme, termed the DRL-MCTS algorithm, proves significantly more effective than either the DRL or MCTS alone. Specifically, DRL-MCTS achieves significantly improved performance {\it and} reduced complexity as compared to MCTS, due to a more effective tree search process and a smaller number of playouts required to achieve satisfactory performance. DRL-MCTS outperforms DRL by a large margin in terms of the detection performance, albeit DRL has a lower complexity. DRL-MCTS also outperforms DetNet \cite{Samuel2019} in the detection performance of binary phase shift keying (BPSK) modulation.
\item By using DRL and a reward system to calculate losses instead of using labels from the transmitted symbol vectors, we avoid the previously mentioned labeling dilemma during DNN training. Only the received symbol vector, channel information, and an estimate of the transmitted symbol vector are required in the reward system.
\end{enumerate}

The rest of the paper is organized as follows. Sec.~\ref{sec:system} formulates the MIMO detection problem. Sec.~\ref{sec:MCTS} introduces the MCTS algorithm for MIMO detection. Sec.~\ref{sec:DRL-MCTS} presents the proposed DRL-aided MCTS algorithm. Sec.~\ref{sec:results} presents the simulation results and discussion. Finally, Sec.~\ref{sec:conclusion} concludes the paper. 

{\it Notations:} In this paper, we denote the complex Gaussian distribution with mean $\mu$ and variance $\sigma^2$ as $\mathcal{C}\mathcal{N}(\mu,\,\sigma^{2})$. Boldface uppercase letters (e.g., $\mathbf{H}$) are used to denote matrices, boldface lowercase letters (e.g., $\mathbf{s}$) are used to denote vectors, and lowercase letters (e.g., \textit{s}) are used to denote scalars. $(\cdot)^*$, $(\cdot)^{\top}$, and $(\cdot)^{\dagger}$ denote complex conjugate, transpose, and conjugate transpose, respectively. $\lVert \cdot \rVert$ denotes the $l_2$-norm of a vector. $\Re( \cdot )$ and $\Im( \cdot )$ denote the real and imaginary parts of a complex number, respectively. $\vert \cdot \vert$ denotes the cardinality of a set.

\section{System Model and Problem Statement} \label{sec:system}

We consider an uncoded $N_T\times N_R$ MIMO system with $N_T$ transmit antennas and $N_R$ receive antennas. The complex baseband received signal can be expressed as
\begin{align} \label{eq:Complex_MIMO_eq}
{\mathbf y}_c = {\mathbf H}_c \widetilde{\mathbf{x}}_c + {\mathbf w}_c.
\end{align}
Here, $\widetilde{\mathbf{x}}_c$ is the $N_T\times1$ transmit signal vector containing uncorrelated entries selected equiprobably from a square quadrature amplitude modulation (QAM) alphabet $\mathcal{S} = \{a + ib \: | \: a, b \in \mathcal{Q} \}$, where $\mathcal{Q}$ is the pulse amplitude modulation (PAM) alphabet. $\widetilde{\mathbf{x}}_c$ has zero mean and covariance matrix ${\sigma_{x}^2}\mathbf{I}_{N_T}$. ${\mathbf H}_c$ is the $N_R\times N_T$ channel matrix containing independent and identically distributed (i.i.d.) complex Gaussian elements with zero mean and unit variance, and is assumed perfectly known at the receiver but not at the transmitter. ${\mathbf w}_c$ is the additive white Gaussian noise (AWGN) with i.i.d. complex elements and has zero mean and covariance matrix ${\sigma_{w}^2}\mathbf{I}_{N_R}$. The complex signal model \eqref{eq:Complex_MIMO_eq} can be transformed to an equivalent real signal model by defining $\mathbf{y'} = \left[ \Re({\mathbf y}_c) \: \Im({\mathbf y}_c) \right]^{\top}$, $\mathbf{\widetilde{x}} = \left[ \Re(\widetilde{\mathbf{x}}_c) \: \Im(\widetilde{\mathbf{x}}_c) \right]^{\top}$, $\mathbf{w} = \left[ \Re({\mathbf w}_c) \: \Im({\mathbf w}_c) \right]^{\top}$, and $\mathbf{H}=\begin{bsmallmatrix} \Re({\mathbf H}_c) & -\Im({\mathbf H}_c)\\ \Im({\mathbf H}_c) & \Re({\mathbf H}_c) \end{bsmallmatrix}$. The resulting real-valued model is given by
\begin{align} \label{eq:Real_MIMO_eq}
\mathbf{y}' = \mathbf{H}\widetilde{\mathbf{x}} + \mathbf{w}
\end{align}
where $\mathbf{y}' \in \mathbb{R}^{n}$, $\mathbf{H} \in \mathbb{R}^{n \times m}$, $\mathbf{\widetilde{x}} \in \mathcal{Q}^{m}$, and $\mathbf{w} \in \mathbb{R}^{n}$, with $n = 2N_R$ and $m = 2N_T$. 

It is well-known that the optimal maximum likelihood (ML) detector to recover the transmitted symbol vector $\mathbf{\widetilde{x}}$ from $\mathbf{y}'$ is to find $\widetilde{\mathbf{x}}_{\rm ML} = \mathop{\arg\min}_{\mathbf{x} \in \mathcal{Q}^m } \|  \mathbf{y'} - \mathbf{Hx} \|^2$. By a QR decomposition of the channel matrix $\mathbf{H}$ such that $\mathbf{H} = \mathbf{QR}$, where $\mathbf{Q} \in \mathbb{R}^{n \times m}$ is an orthogonal matrix and $\mathbf{R} \in \mathbb{R}^{m \times m}$ is an upper-triangular matrix, the ML detection criterion can equivalently be expressed as $\widetilde{\mathbf{x}}_{\rm ML} = \mathop{\arg\min}_{\mathbf{x} \in \mathcal{Q}^m} \| \mathbf{y}-\mathbf{Rx} \|^2$, where $\mathbf{y} = \mathbf{Q}^{\top}\mathbf{y}'$. Due to the upper-triangular structure of $\mathbf{R}$, $\| \mathbf{y}-\mathbf{Rx} \|^2$ can be expanded as the summation of $m$ terms, where the $(m-k+1)$th term (for $k=1, 2, \ldots, m$) depends only on the partial symbol vector $\mathbf{x}_{k}^m\triangleq (x_k, \ldots, x_m)^{\top} \in \mathcal{Q}^{m-k+1}$:
\begin{align} \label{eq:expanded_ML_eq}
(y_m - r_{m,m}x_m)^2 &+ \left(y_{m-1} - \sum_{i=m-1}^{m} r_{m-1,i}x_i\right)^2 + \cdots \nonumber\\
&+ \left(y_1 - \sum_{i=1}^{m} r_{1,i}x_i\right)^2
\end{align}
where $y_i$ is the $i$th element of $\mathbf{y}$, $x_i$ is the $i$th element of $\mathbf{x}$, and $r_{i,j}$ is the $(i,j)$-entry of $\mathbf{R}$. We denote the $(m-k+1)$th term in \eqref{eq:expanded_ML_eq} by $b(\mathbf{x}_{k}^m)$ and the summation of the first $m-k+1$ terms by $d(\mathbf{x}_{k}^m)$. Then, ML detection becomes 
\begin{align} \label{eq:ML_eq}
\widetilde{\mathbf{x}}_{\rm ML} = \mathop{\arg\min}_{{\mathbf x}_1^m \in \mathcal{Q}^m } \ d(\mathbf{x}_{1}^m). 
\end{align}
Note that \eqref{eq:expanded_ML_eq} creates a rooted tree structure that allows sequentially determining the transmitted signal vector $\mathbf{x}$ from $\mathbf{x}_{m}^m$ to $\mathbf{x}_{1}^m$ by detecting one element of $\mathbf{x}$ at a time. Specifically, we can first recover $x_{m}$ and calculate $b(\mathbf{x}_{m}^m)$, and then recover $x_{m-1}$ with known $x_{m}$ to obtain $\mathbf{x}_{m-1}^m$ and calculate $b(\mathbf{x}_{m-1}^m)$, etc.

\begin{figure*}[tb!]
    \centering
    \subfigure[]{
        \centering
        \includegraphics[width={\linewidth}]{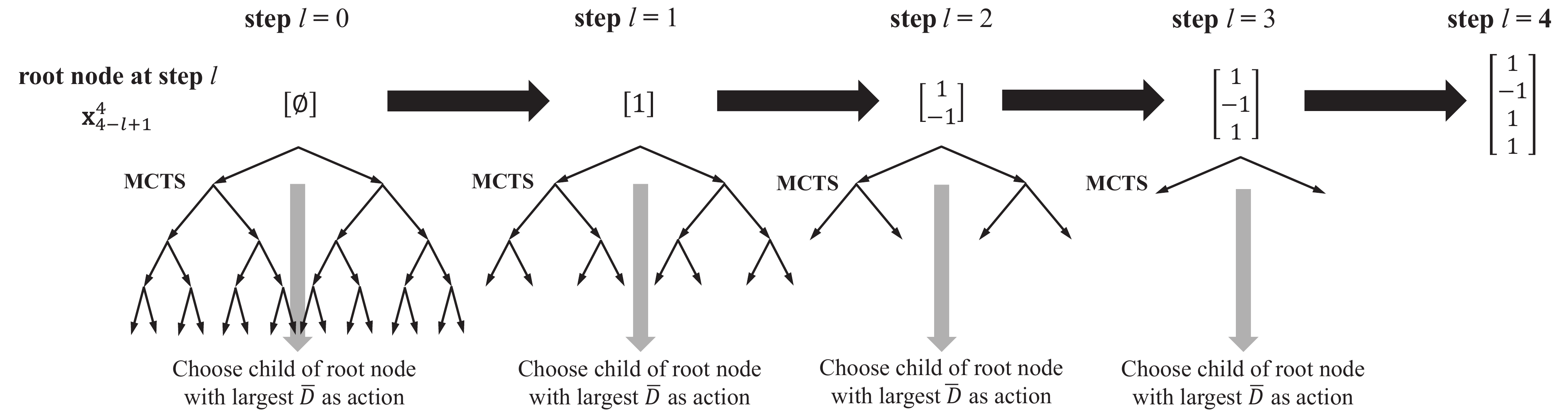}
        \label{fig:MCTS_a}}
    \subfigure[]{
        \centering
        \includegraphics[width={\linewidth}]{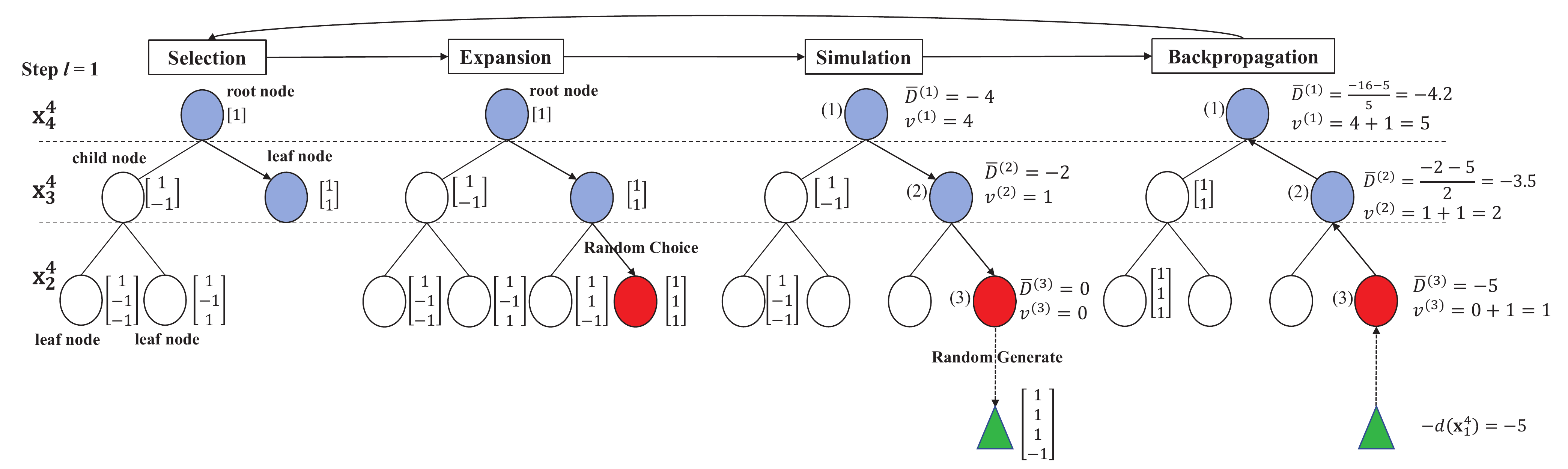}
        \label{fig:MCTS_b}}
    \caption{MCTS for MIMO detection, for the example of a $2 \times 2$ QPSK system ($m=4$ and $|{\cal Q}|=2$). (a) MCTS step progression. A total of four steps are needed to finish detection. In each step, MCTS is performed to recover an additional transmit symbol element until the terminal state is reached. (b) MCTS algorithm at step $l=1$. Each node in the tree represents a unique partial symbol vector. In Selection, a leaf node is reached using UCT. During Expansion, all possible child nodes are added to the tree and become the new leaf nodes. In Simulation and Backpropagation, where the node index $i$ is shown to the left side of each node in parentheses and the node statistics $\overline{D}^{(i)}$ and $v^{(i)}$ are shown to the right side of each node, node statistics are updated by backpropagating the simulated $-d(\mathbf{x}_{1}^4)$ result.}
    \label{fig:MCTS}
\end{figure*}

\section{Monte Carlo Tree Search (MCTS) Algorithm} \label{sec:MCTS}

The MCTS algorithm has found applications in many discrete decision making problems, such as chess and video games, where MCTS tries to output a best action in a discrete action space based on the current state. A tree search is performed each time we wish to take a new action. In the case of MIMO detection, an action is defined as recovering an element of the transmitted symbol ${\mathbf x}$, and the action space is the PAM alphabet $\mathcal{Q}$ with $\vert \mathcal{Q} \vert$ discrete actions. A terminal state is reached when a complete symbol vector $\mathbf{x}_{1}^m={\mathbf x}$ is recovered. A total of $m$ actions/tree searches are required to get to a terminal state. 

MCTS differs from other tree search algorithms for MIMO detection in that the tree is actually constructed along the search process. At the start of each tree search, the Monte Carlo tree consists only of the root node, and as the search progresses, the tree expands by adding new leaf nodes to the existing tree. The root node of the search tree is the current state at some step $l$ ($l = 0,1,2,\ldots,m-1$) and represents the partial symbol vector $\mathbf{x}_{m-l+1}^m$. Note that at step $l=0$, the root node is a virtual root node as we define $\mathbf{x}_{m+1}^m \triangleq \emptyset$; at step $l=m$, the root node is a nominal root node and the detection steps are completed. The algorithm aims to determine the best $x_{m-l}$ at step $l$ so that at the end of detection \eqref{eq:expanded_ML_eq} is minimized. Fig.~\ref{fig:MCTS_a} shows the MCTS step progression, for the example of a $2 \times 2$ QPSK system ($m=4$ and $|{\cal Q}|=2$). Fig.~\ref{fig:MCTS_b} shows the MCTS algorithm in search of the best action $x_{m-l}$ at each step $l$ (here, $l=1$ is shown), as further discussed below. 

Each node $i$ in the tree is associated with two node statistics: 
\begin{enumerate}
\item[i)] the expected $-d(\mathbf{x}_{1}^m)$ that this node will achieve by going downward through layers of tree from this node, which is denoted as $\overline{D}^{(i)}$; and 
\item[ii)] the number of times this node has been selected during the MCTS process, which is denoted as $v^{(i)}$. 
\end{enumerate}
Note that a negative sign is added in front of $d(\mathbf{x}_{1}^m)$ to convert a metric minimization problem in \eqref{eq:ML_eq} into a reward maximization problem in the framework of MCTS. $\overline{D}^{(i)}$ and $v^{(i)}$ are initially set to zero.

The MCTS algorithm comprises four steps:
\begin{enumerate}
\item {\it Selection:} Starting from the root node, successive child nodes are selected until a leaf node is reached. Specifically, a child node that maximizes the upper confidence bound for trees (UCT) \cite{cameron2012mcts}, defined as
\begin{align} \label{eq:uct}
\overline{D}^{(i)} + c_{\rm uct}\sqrt{\frac{\ln{V^{(i)}}}{v^{(i)}}},
\end{align}
is selected. Here, $V^{(i)}$ is the number of times the parent node of a node $i$ has been selected, and $c_{\rm uct}$ is a positive-valued exploration parameter which is empirically determined. The first term of UCT enforces a reward-based selection, and the second term of UCT encourages exploration since a smaller $v^{(i)}$ leads to a higher value of UCT and therefore a higher chance of selection.
\item {\it Expansion:} If the selected leaf node is not a terminal state, $\vert \mathcal{Q} \vert$ child nodes are added to the tree. If the selected leaf node is a terminal state, no new leaf node will be added, the Simulation step will be skipped, and the Backpropagation step will be performed.
\item {\it Simulation:} Randomly select one of the child nodes that has just been added to the tree in the Expansion step, and simulate detection by expanding the partial symbol vector with random elements $x \in \mathcal{Q}$ until $\mathbf{x}_{1}^m$ is reached.
\item {\it Backpropagation:} The mock detection result $\mathbf{x}_{1}^m$ from either the Simulation or Expansion steps is evaluated by calculating $-d(\mathbf{x}_{1}^m)$. The value is fed back upward toward the root node of the tree to update node statistics of the leaf node and all the nodes selected during the MCTS process. Specifically, $\overline{D}^{(i)}$ is updated by averaging all $-d(\mathbf{x}_{1}^m)$ values backpropagated from the leaf node below node $i$, and $v^{(i)}$ is updated by adding by one if node $i$ is on the backpropagation path. 
\end{enumerate}
A completion of these four procedures is termed one {\it playout} and the total number of playouts is a tunable parameter with performance-complexity tradeoffs. After completing all playouts from the current state, the action represented by the best child node (corresponding to the largest $\overline{D}^{(i)}$) among all child nodes of the root node is selected as the best action $x_{m-l}$ at step $l$.

To increase the search efficiency, we also added two tips to the MIMO MCTS algorithm:
\begin{enumerate}
\item Instead of re-growing a whole tree from a root node after taking a best action each step $l$ for $l>0$, we  make the best child node after all playouts the new root node and preserve all node statistics of the subtree spanning from the best child node. By reusing node statistics from previous steps, we build on the cumulative experiences which can lead to more accurate final reward estimations.
\item By reusing the node statistics of previous trees, we can actually decrease the number of playouts at latter steps to increase detection speed. To achieve this, we take the floor function of the playout number multiplied with a decay factor $\beta_{p} \in (0 , 1)$ after each step.
\end{enumerate}

As can be seen, the general MCTS algorithm relies on {\it random} sampling in the Simulation step whose outcomes are then used to approximate the expected $-d(\mathbf{x}_{1}^m)$. While UCT allows sampling paths efficiently by growing asymmetrical trees and focusing more on promising subtrees, good solutions could still be overlooked if the sample size is not large enough. This motivates an enhancement of MCTS with a learned and guided Simulation step enabled by DRL, so that a better decision could be achieved. This is presented next.

\section{Deep Reinforcement Learning Aided Monte Carlo Tree Search (DRL-MCTS) Algorithm} \label{sec:DRL-MCTS}

We propose to incorporate a DRL architecture into the MCTS framework. Our DRL architecture, inspired by DeepMind's AlphaGo Zero \cite{David2017alphago}, consists of a policy value network and a state value network, both taking some defined state $\mathbf{s}_l$ at step $l$ of the detection process as input, and outputting the \textit{policy value} $\widehat{\mathbf{p}}_l$ and the \textit{state value} $u_l$, respectively. Here, $\widehat{\mathbf{p}}_l$ has $\vert \mathcal{Q} \vert$ elements, representing the probability distribution over $\vert \mathcal{Q} \vert$ discrete possible moves, and $u_l$ represents the final expected $-d(\mathbf{x}_{1}^m)$ of that state $s_l$. The two outputs are integrated into the MCTS algorithm, where the policy value is adopted into the UCT for enhanced exploration, and the state value is directly used in place of the Simulation step as the expected final outcome.

\subsection{Definition of State, Action, and Reward}

We first define the state, action, and reward for the MIMO detection problem under a reinforcement learning framework:
\begin{itemize}
\item {\it State:} The state vector $\mathbf{s}_l$ observed by the agent at step $l$ ($l=0,\ldots, m-1$) consists of some compressed sufficient representations of the channel matrix ${\mathbf H}$, the received signal ${\mathbf y}'$, the partial symbol vector $\mathbf{x}_{m-l+1}^m$ and its metrics, i.e., 
\begin{align} \label{eq:input_state}
\mathbf{s}_l = \begin{bmatrix}
\mathbf{y}\\\mathbf{y}'\\\mathbf{H}^{\top}\mathbf{y}'\\\boldsymbol{\chi}_{m-l+1}^m\\\boldsymbol{\chi}_{m-l+2}^m\\ b(\mathbf{x}_{m-l+1}^m)\\ d(\mathbf{x}_{m-l+1}^m)
\end{bmatrix}
\end{align}
where $\boldsymbol{\chi}_{m-l+1}^m = [\mathbf{0}_{1\times (m-l)}, (\mathbf{x}_{m-l+1}^m)^{\top}]^{\top}$ and $\boldsymbol{\chi}_{m-l+2}^m$ (similarly defined) are the zero-padded partial symbol vectors at step $l$ and step $l-1$, respectively, both having a fixed dimension $m \times 1$. Note that at step $l=0$, the last four entries of ${\mathbf s}_l$ degenerate to zero vectors/scalars as we define $\boldsymbol{\chi}_{m+1}^m \triangleq \mathbf{0}_{m \times 1}$, $\boldsymbol{\chi}_{m+2}^m \triangleq \mathbf{0}_{m \times 1}$, $b(\mathbf{x}_{m+1}^m) \triangleq 0$, and $d(\mathbf{x}_{m+1}^m) \triangleq 0$; at step $l=1$, $\boldsymbol{\chi}_{m-l+2}^m = \boldsymbol{\chi}_{m+1}^m = \mathbf{0}_{m \times 1}$.
\item {\it Action:} The action $a_l$ at step $l$ recovers $x_{m-l}$ with $\vert \mathcal{Q} \vert$ possible actions.
\item {\it Reward:} The reward value $r_l$ for an action $a_l$ at step $l$ is $-d(\mathbf{x}_{m-l}^m)$. Note that the reward for DRL slightly differs from the Simulation step in the MCTS algorithm, where only $-d(\mathbf{x}_{1}^m)$ of the full recovered symbol vector is used to update the node statistics. 
\end{itemize}

The training data for the policy value network and state value network are collected by sending $N$ transmit symbol vectors through a fixed channel and recording the detection process at the receiver with the tuples $(s_{l}^{j}, a_{l}^{j}, r_{l}^{j}, s_{l+1}^{j})$, where the superscript denotes the $j$th transmit symbol vector ($j = 1, 2, \ldots, N$). These collected data are used to update the networks, which in turn generate new sets of data. The process repeats until the networks converge. 

Note that unlike the case of playing the game of Go where each action taken does not always return an immediate reward, in MIMO detection, after performing an action we can calculate the reward to evaluate the action immediately. Thus, the network training can be more efficient since each action is evaluated independently.

\begin{figure*}[tb!]
    \centering
    \subfigure[]{
        \centering
        \includegraphics[width=0.95\columnwidth]{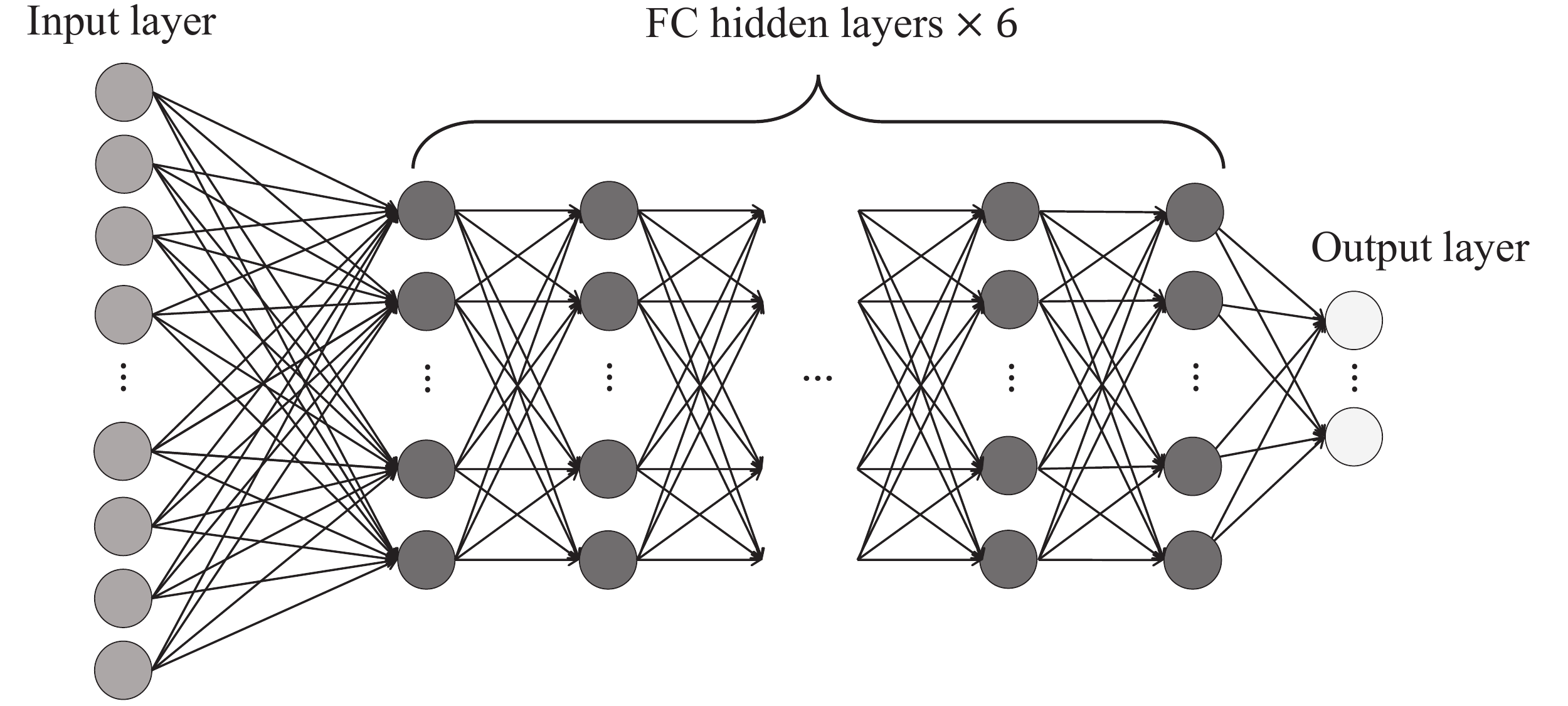}
        \label{fig:network_topology_a}}
    \subfigure[]{
        \centering
        \includegraphics[width=0.95\columnwidth]{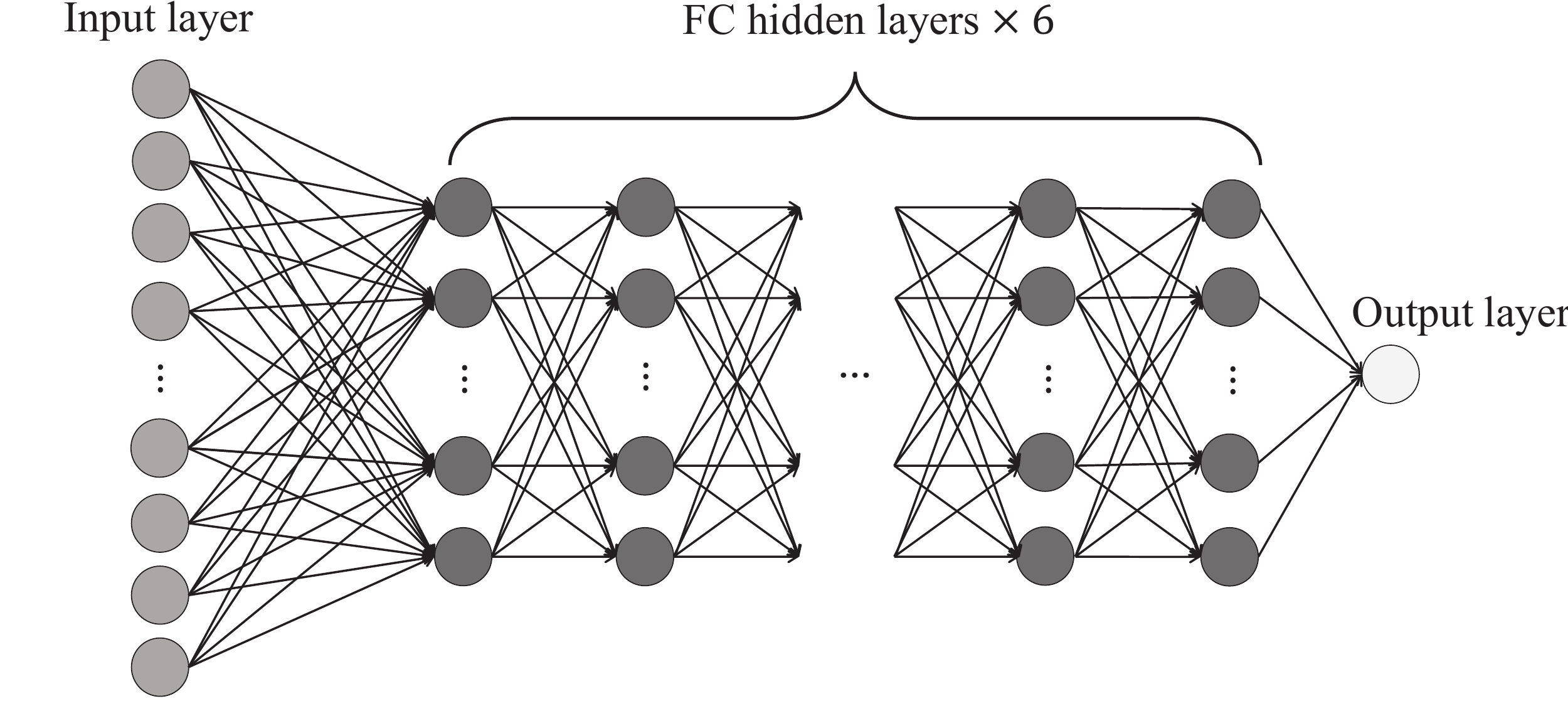}
        \label{fig:network_topology_b}}
    \caption{Network architecture of neural networks used in DRL. (a) The actor of the policy value network. (b) The critic of the policy value network, and the state value network. The detailed configurations of each network are given in Table~\ref{table:network config}.}
    \label{fig:network_topology}
\end{figure*}

\begin{table*}[ht]
\centering
\caption{Network Configurations of Neural Networks Used in DRL}
\begin{tabular}{|c|c|c|c|c|}
\hline
\textbf{Network} & \textbf{Input Layer Size}& \textbf{Hidden Layer Size}& \textbf{Output Layer Size} & \textbf{Activation Function}                      \\ \hline
Policy Value (Actor)   & $4m+n+2$  & 
\begin{tabular}[c]{@{}c@{}}$4m+n+2$, $8N_T$, $8N_T$, $8N_T$, $8N_T$, $8N_T$\end{tabular} 
& $\vert \mathcal{Q} \vert$ & 
\begin{tabular}[c]{@{}c@{}}ReLU, ReLU, ReLU, ReLU, \\ ReLU, ReLU, softmax\end{tabular} \\ \hline
Policy Value (Critic)  & $4m+n+2$  & 
\begin{tabular}[c]{@{}c@{}}$4m+n+2$, $8N_T$, $8N_T$, $8N_T$, $8N_T$, $8N_T$ \end{tabular} 
& 1 & 
\begin{tabular}[c]{@{}c@{}}tanh, tanh, tanh, tanh, tanh, tanh\end{tabular}          \\ \hline
State Value & $4m+n+2+\vert \mathcal{Q} \vert$ & \begin{tabular}[c]{@{}c@{}}$4m+n+2$, $8N_T$, $8N_T$, $8N_T$, $8N_T$, $8N_T$\end{tabular} 
& 1 & 
\begin{tabular}[c]{@{}c@{}}tanh, tanh, tanh, tanh, tanh, tanh\end{tabular}          \\ \hline
\end{tabular}
\label{table:network config}
\end{table*}

\subsection{Policy Value Network}

As mentioned earlier, the policy value network outputs a vector $\widehat{\mathbf{p}}_l$ representing the probability distribution over $\vert \mathcal{Q} \vert$ actions in the action space. Our policy value network uses an actor-critic architecture \cite{konda2000,volodymyr2016} consisting of two fully-connected neural networks. The actor network is composed of an input layer, 6 fully-connected (FC) hidden layers, and a final output layer that outputs $\widehat{\mathbf{p}}_l$, as shown in Fig.~\ref{fig:network_topology_a}. The critic network, as shown in Fig.~\ref{fig:network_topology_b}, has almost the same network structure but outputs a scalar value $q_l$ representing the expected discounted cumulative reward, or Q-value, from the current input state until the terminal state. Note that $q_l$ is only used during the training process to train the actor network and will not be used in the final detection process. The loss function of the critic network incorporates the squared temporal difference (TD) error as:
\begin{align} \label{eq:Loss_critic}
L(\boldsymbol{\theta}_c)=\frac{1}{N}\sum_{j=1}^{N}\sum_{l=0}^{m-1}(r_{l}^{j}+\gamma^{m-l} q_{l+1}^{j}-q_{l}^{j})^2 + c_1 \lVert \boldsymbol{\theta}_c \rVert^2
\end{align}
where the TD error is defined as:
\begin{align} \label{eq:TD_error}
\mbox{TD} \triangleq r_{l}^{j}+\gamma^{m-l}q_{l+1}^{j}-q_{l}^{j}.
\end{align}
In \eqref{eq:Loss_critic}, $\boldsymbol{\theta}_c$ represents the critic network parameters; $\gamma \in [0.9, 1]$ is a discount value; $q_{l}^{j}$ and $q_{l+1}^{j}$ are the critic network outputs for input states $\mathbf{s}_l$ and $\mathbf{s}_{l+1}$, respectively, for the $j$th transmit symbol vector; $c_1$ is a scaling constant; and $\lVert \boldsymbol{\theta}_c \rVert^2$ is the $l2$ regularization of the network parameters. Note that $q_m^{j} \triangleq 0$, since after step $l=m-1$, the detection terminates and there is no expected reward.   

The loss function of the actor network is defined as
\begin{align} \label{eq:Loss_actor}
L(\boldsymbol{\theta}_a)=\frac{1}{N}\sum_{j=1}^{N} & \sum_{l=0}^{m-1} \big(\mbox{TD}\times \mathcal{L}_{\rm CE}(\mathbf{p}_l^{j}, \widehat{\mathbf{p}}_l^{j}) \nonumber\\
& -c_2 \times \mathcal{L}_{\rm CE}(\widehat{\mathbf{p}}_l^{j}, \widehat{\mathbf{p}}_l^{j})\big) + c_3 \lVert \boldsymbol{\theta}_a \rVert^2
\end{align}
where $\mathcal{L}_{\rm CE}(\cdot, \cdot)$ is the cross-entropy function:
\begin{align}\label{eq:cross_entropy}
\mathcal{L}_{\rm CE}(\mathbf{p}_l^{j}, \widehat{\mathbf{p}}_l^{j}) = -\sum_{k=1}^{\vert \mathcal{Q} \vert}\mathbf{p}_{l,k}^{j}\log \widehat{\mathbf{p}}_{l,k}^{j}
\end{align}
and $\boldsymbol{\theta}_a$ is the actor network parameters. $\mathbf{p}_{l}^{j}$ represents the true probability of action taken in step $l$ for the $j$th transmit symbol vector. In fact, $\mathbf{p}_{l}^{j}$ is the one-hot encoding of the element $x_{m-l}$ recovered at step $l$. $\widehat{\mathbf{p}}_{l}^{j}$ is the predicted probability distribution from the actor network. The subscript $k$ in \eqref{eq:cross_entropy} denotes the $k$th element in $\mathbf{p}_{l}^{j}$ and $\widehat{\mathbf{p}}_{l}^{j}$. The term $\mathcal{L}_{\rm CE}(\widehat{\mathbf{p}}_l^{j},\widehat{\mathbf{p}}_l^{j})$ in \eqref{eq:Loss_actor} is the entropy of $\widehat{\mathbf{p}}_{l}^{j}$, which is used to keep the output probability distribution more even, leading to better exploration as we sample actions from the distribution during training. $c_2$ and $c_3$ are both scaling constants. The last term in \eqref{eq:Loss_actor} is the $l2$ regularization of the network parameters. This actor-critic network architecture allows us to train a strong actor to output good policy values to increase the efficiency and accuracy of the MCTS algorithm for MIMO detection. 

\subsection{State Value Network}

As mentioned earlier, the state value network outputs a scalar $u_l$ as the predicted $-d(\mathbf{x}_{1}^m)$ to replace the Simulation step in the original MCTS algorithm. The state value network structure is shown in Fig.~\ref{fig:network_topology_b}. The input to the state value network at step $l$ is ${\mathbf s}_l$ in \eqref{eq:input_state} concatenated with the policy value network output $\widehat{\mathbf{p}}_l$, i.e., $\mathbf{s}'_l = [\mathbf{s}_l^{\top}, \widehat{\mathbf{p}}_l^{\top}]^{\top}$. The loss function of the state value network is defined as
\begin{align} \label{eq:Loss_state_value}
L(\boldsymbol{\theta}_s)=\frac{1}{N}\sum_{j=1}^{N}\sum_{l=0}^{m-1}(u_l^{j}-D^{j})^2+c_4 \lVert \boldsymbol{\theta}_s \rVert^2
\end{align}
where $\boldsymbol{\theta}_s$ is the network parameters, $u_l^{j}$ is the state value network output at step $l$ for the $j$th transmit symbol vector, $D^{j}$ is the final $-d(\mathbf{x}_1^m)$ for the $j$th transmit symbol vector and is obtained after step $l-1$, $c_4$ is a scaling constant, and $\lVert \boldsymbol{\theta}_s \rVert^2$ is the $l2$ regularization of the network parameters. 

Network architecture and configurations of all three networks are shown in Fig.~\ref{fig:network_topology} and Table~\ref{table:network config}. The network configurations are determined empirically and the optimization of the network sizes can be further explored in future work. 

\begin{figure}[tb!]
    \centering
    \includegraphics[width=0.95\columnwidth]{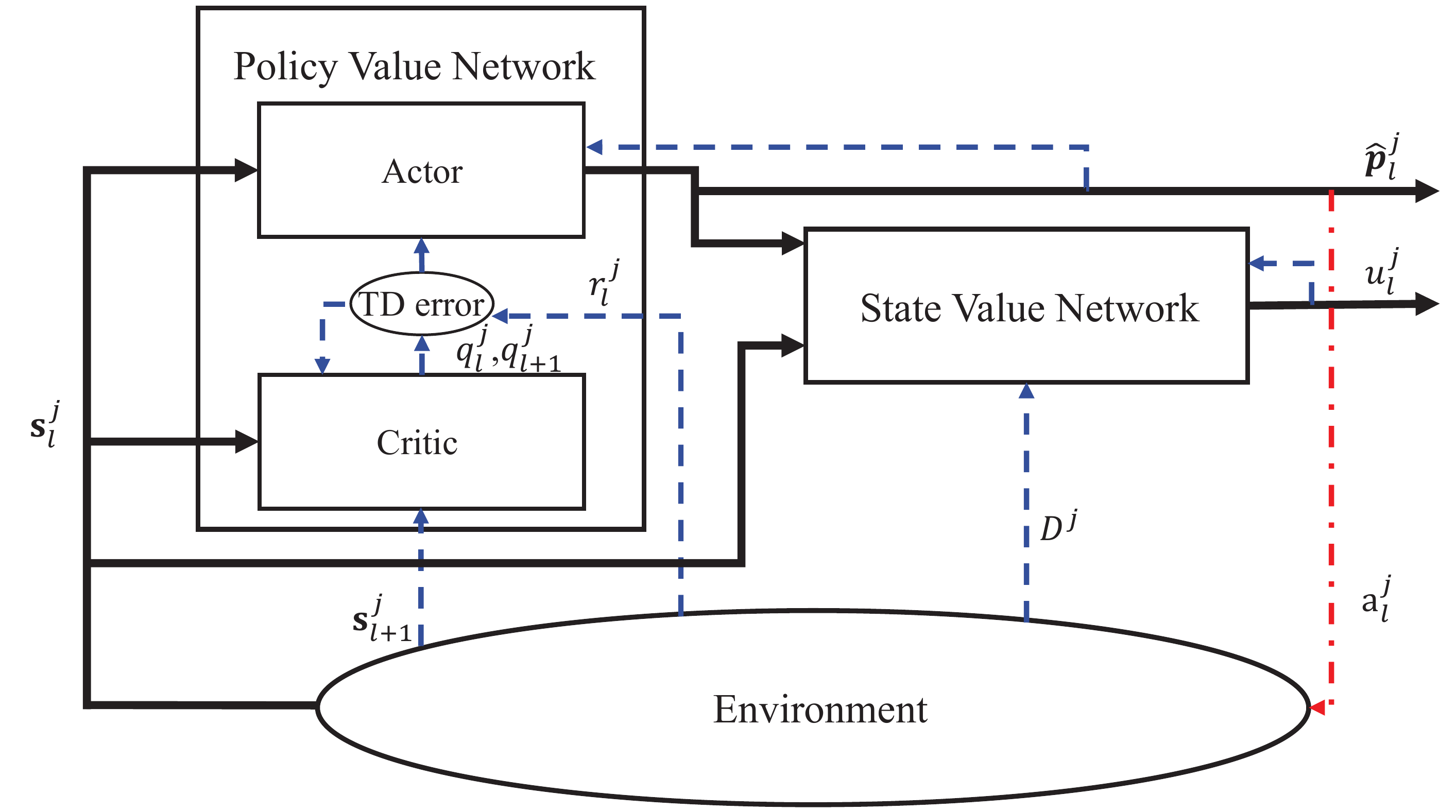}
    \caption{Summary of the proposed DRL architecture. The blue dashed lines indicate values used to calculate losses for the networks. The red dashed-dotted line indicates the action $a_l^{j}$ sampled from the policy value network output $\widehat{\mathbf{p}}_l^{j}$ at step $l$ for the $j$th transmit symbol vector.}
    \label{fig:RL_framework}
\end{figure}

The proposed DRL architecture is summarized in Fig.~\ref{fig:RL_framework}.

\subsection{The DRL-MCTS Algorithm}

\begin{figure*}[tb!]
    \centering
    \includegraphics[width={0.8\linewidth}]{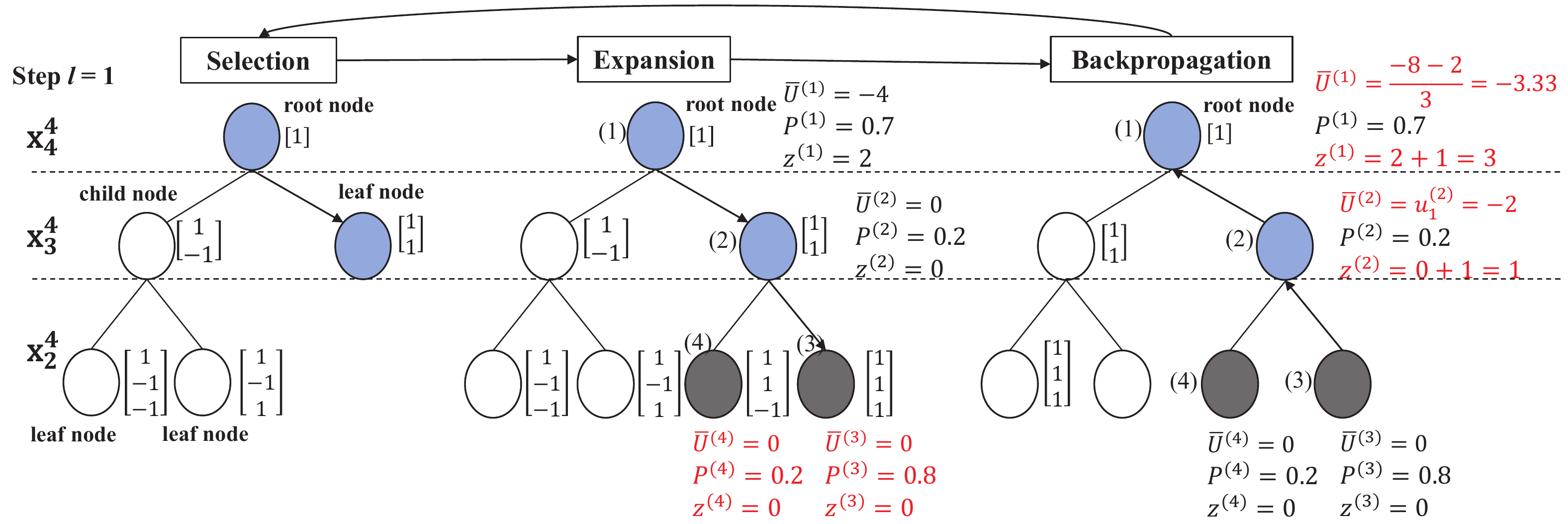}
    \caption{(Cf. Fig.~\ref{fig:MCTS_b}, for the same example of a $2 \times 2$ QPSK system) DRL-MCTS algorithm at step $l=1$. In Selection, a leaf node is reached using PUCT. During Expansion, all possible child nodes are added to the tree and become the new leaf nodes. In Expansion and Backpropagation, where the node index $i$ is shown to the left side of each node in parentheses and the node value $\overline{U}^{(i)}$, prior node probability $P^{(i)}$, and selection count $z^{(i)}$ are shown to the right side of each node, child nodes are added to the tree (Expansion) and node statistics are updated and backpropagated (Backpropagation).}
    \label{fig:MCTS_RL}
\end{figure*}

By a slight abuse of notation, we use $\widehat{\mathbf{p}}_l^{(i)}$ and $u_l^{(i)}$ to denote the outputs of the policy value network and state value network, respectively, corresponding to the transmit symbol vector represented by node $i$ in the tree at step $l$. Each node $i$ in the tree is associated with three node statistics:
\begin{enumerate} 
\item[i)] the expected $-d(\mathbf{x}_{1}^m)$, or \textit{node value}, that this node $i$ will achieve as evaluated by the state value network output, which is denoted as $\overline{U}^{(i)}$; 
\item[ii)] a \textit{prior node probability}, $P^{(i)}$, which is a probability taken directly from one of the elements in the policy value network output of the parent node of node $i$; and 
\item[iii)] the number of times this node $i$ has been selected, or the \textit{selection count}, during the DRL-MCTS process, which is denoted as $z^{(i)}$. 
\end{enumerate}
$\overline{U}^{(i)}$ and $z^{(i)}$ are initialized to $0$, whereas $P^{(i)}$ is initialized directly from the policy value network output of the parent node of node $i$. 

The DRL-MCTS algorithm comprises three steps:
\begin{enumerate}
\item {\it Selection:} The process is the same as the Selection step in the MCTS algorithm, except that a variant of UCT, defined as
\begin{align} \label{eq:puct}
\overline{U}^{(i)} + c_{\rm puct}P^{(i)}\frac{\sqrt{Z^{(i)}}}{1+z^{(i)}},
\end{align}
is used for node selection \cite{David2017alphago}. Here, $Z^{(i)}$ is  the  number of times the parent node of node $i$ has been selected.
\item {\it Expansion:} If the selected leaf node $i$ is not a terminal state, $\widehat{\mathbf{p}}_l^{(i)}$ and $u_l^{(i)}$ are evaluated by the trained neural networks, where the $\vert \mathcal{Q} \vert$ values in $\widehat{\mathbf{p}}_l^{(i)}$ are used to initialize the prior node probabilities of $\vert \mathcal{Q} \vert$ child nodes as the child nodes are added to the tree. If the selected leaf node is a terminal state, no new leaf node will be added, and Backpropagation will be performed. 
\item {\it Backpropagation:} If the leaf node $i$ is not a terminal state, $u_l^{(i)}$ is backpropagated, where the node value of each node in the Selection path is updated by averaging the leaf node value $u_l^{(i)}$ with previous recorded node values of that node, and $z^{(i)}$ is updated by adding by one if node $i$ is on the backpropagation path. If the leaf node $i$ is a terminal state, directly compute $-d(\mathbf{x}_1^m)$ and the value is backpropagated toward the root node of the tree to update node values and selection counts of all the nodes selected during the Selection step.
\end{enumerate}
After completing a designated number of playouts from the current state, the action represented by the best explored child node (corresponding to the largest $z^{(i)}$) among all child nodes of the root node is selected as the best action $x_{m-l}$ at step $l$. An example of the DRL-MCTS algorithm is illustrated in Fig.~\ref{fig:MCTS_RL}. To increase the search efficiency, we also added the two tips mentioned in Sec.~\ref{sec:MCTS}, where we reuse node statistics from previous steps and decay the playout number after each step.

\section{Simulation Results and Discussion} \label{sec:results}

In this section, we analyze the performance and complexity of our proposed method and compare it against other linear and DNN-based detectors under varying channel conditions and different antenna/modulation configurations. We first present the implementation details, followed by results and discussion.

\subsection{Implementation Details} \label{subsec:results-settings}

The proposed method (DRL-MCTS) is implemented using Python codes and the DRL networks are realized using the Tensorflow library. We train the networks and perform the simulations on an Intel i7-8700 CPU processor with NVIDIA GeForce GTX 1070 GPU. To stabilize training and decorrelate the self-play data used to update the DRL networks, threading and multiprocessing libraries are used to create up to $12$ agents, where each agent interacts with its own environment and calculates a local gradient after detecting a complete signal vector using the loss functions defined in Sec.~\ref{sec:DRL-MCTS}. The combined local gradients from all the agents are then used to update the parameters of a single global network, which is then downloaded to each agent to generate new self-play data using the latest parameters. The networks are optimized using the RMSProp optimizer, where the learning rates for the actor, critic, and state value network are all initially set to $0.0001$. The discount value $\gamma$ in \eqref{eq:Loss_critic} is set to $0.95$, and the $l2$ regularization coefficients $c_1$, $c_3$, and $c_4$ are all set to $0.0001$. The scaling constant $c_2$ for the entropy term in \eqref{eq:Loss_actor}, $c_{\rm uct}$, and $c_{\rm puct}$ for the MCTS and DRL-MCTS process will be provided in the discussion for each configuration.

We train our detector under a varying channel model similar to \cite{JeonLee2020}, where an $N_{R}\times N_{T}$ complex channel matrix $\mathbf{H}_c$ is first randomly generated with i.i.d. complex Gaussian elements from $\mathcal{CN}(0,\,1)$ and then the $j$th randomly generated symbol vector $\widetilde{\mathbf{x}}_c^j$ is transmitted through a varying channel:

\begin{align} \label{eq:channel_model}
{\mathbf H}^j_c = \sqrt{1-\epsilon^2}{\mathbf H}_c  + \epsilon\mathbf{W}_c^j,
\end{align}
where $\mathbf{H}^j_c$ is the channel matrix corresponding to the $j$th transmitted symbol vector, $\epsilon \in [0, 1]$ is a constant, and $\mathbf{W}^j_c$ is an $N_{R}\times N_{T}$ complex noise matrix sampled from $\mathcal{C}\mathcal{N}(0,\,1)$ corresponding to the $j$th transmitted symbol vector. Note, however, in the case of BPSK, the channel matrix $\mathbf{H}_c$ and the noise matrices $\mathbf{W}^j_c$ are $N_{R}\times N_{T}$ real matrices with i.i.d. Gaussian elements sampled from $\mathcal{N}(0,\,1)$. Testing data used for simulation results are generated in the same fashion but separately from the training data set. The signal-to-noise ratio (SNR) is defined as $\mathbb{E}\lVert\mathbf{H}_c\widetilde{\mathbf{x}}_c\rVert^2/\mathbb{E}\lVert\mathbf{w}_c\rVert^2  = N_T\sigma_{x}^2/\sigma_{w}^2$, and the symbol error rate (SER) is used to evaluate the performances of the detectors.

\subsection{DRL-MCTS vs. MCTS Comparison} \label{subsec:results-playout}

\begin{figure}[tb!]
    \centering
    \includegraphics[width=\columnwidth]{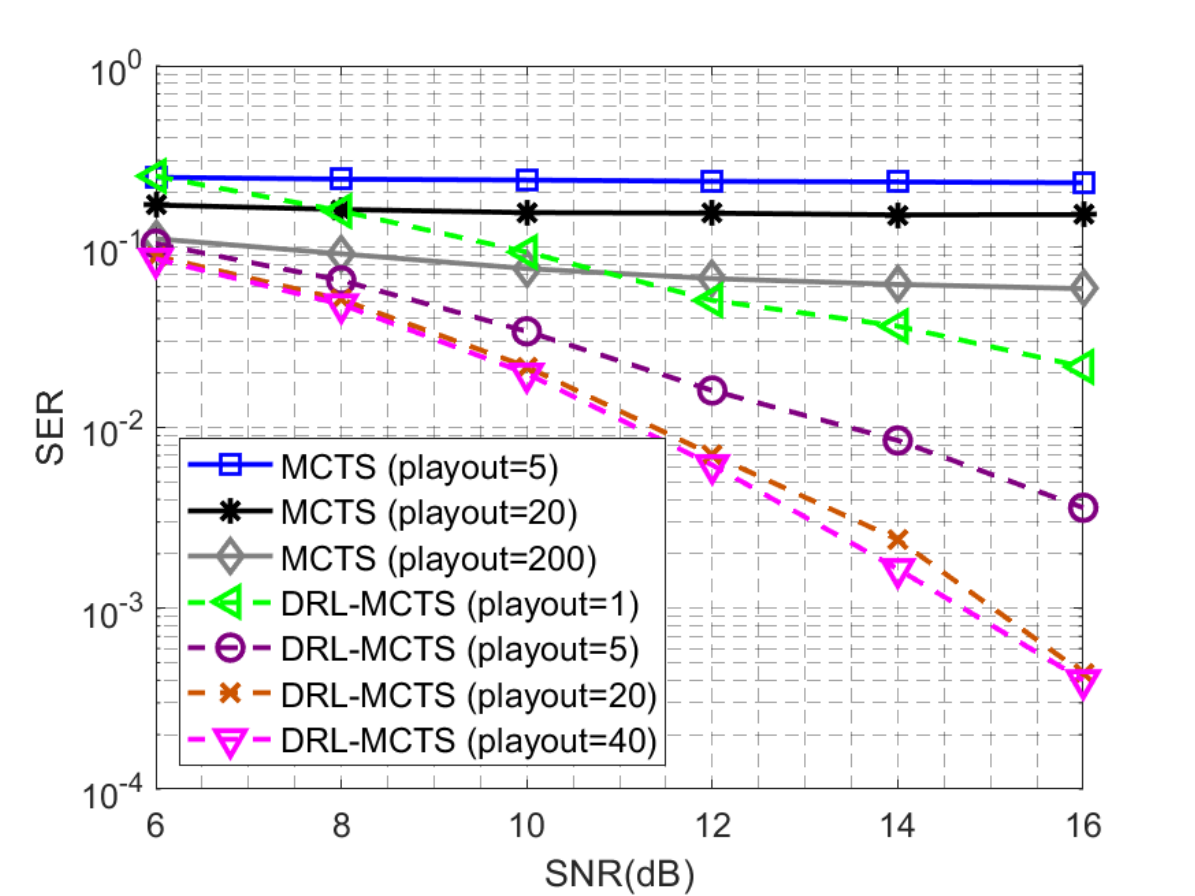}
    \caption{SER performance comparison of MCTS and DRL-MCTS for $8 \times 8$ MIMO with BPSK.}
    \label{fig:playout_8x8BPSK}
\end{figure}

We first compare the proposed DRL-MCTS with specifically MCTS, since DRL-MCTS is based on MCTS and aims to improve its tree search process. Fig.~\ref{fig:playout_8x8BPSK} presents the performance of the MCTS and DRL-MCTS detectors with different playout numbers over an $8\times 8$ real varying MIMO channel model described in \eqref{eq:channel_model} with $\epsilon= 0.1$. 
We set $c_2=1$, $c_{\rm uct}=350$, and $c_{\rm puct}=20$. We adopt playout numbers $5$, $20$, and $200$ for MCTS, and $1$, $5$, $20$, and $40$ for DRL-MCTS. As can be seen, the performance of both detectors improves as the playout number increases. DRL-MCTS outperforms MCTS by a large margin with significantly reduced playout numbers. Since playout number is one of the dominating factors of computational complexity, this suggests that DRL-MCTS demonstrates both performance and complexity advantages as compared to MCTS (more details in Sec.~\ref{subsec:results-complexity}). Note that DRL-MCTS with playout number of one degenerates to detecting with the DRL network directly by taking the action with the largest policy value in $\widehat{\mathbf{p}}_l$ at step $l$, while discarding the state value network output. This setting of DRL-MCTS demonstrates somewhat suffered performance at low SNRs, due to a less reliable policy value network subject to heightened noises. The full capacity of DRL-MCTS is realized when setting the playout number greater than one, where the state value network output is exploited for a more reliable search path by providing the estimated final reward. The figure also shows that a modest number of playouts is sufficient for DRL-MCTS, as additional playouts yield diminishing returns.

\begin{figure}[tb!]
    \centering
    \includegraphics[width=\columnwidth]{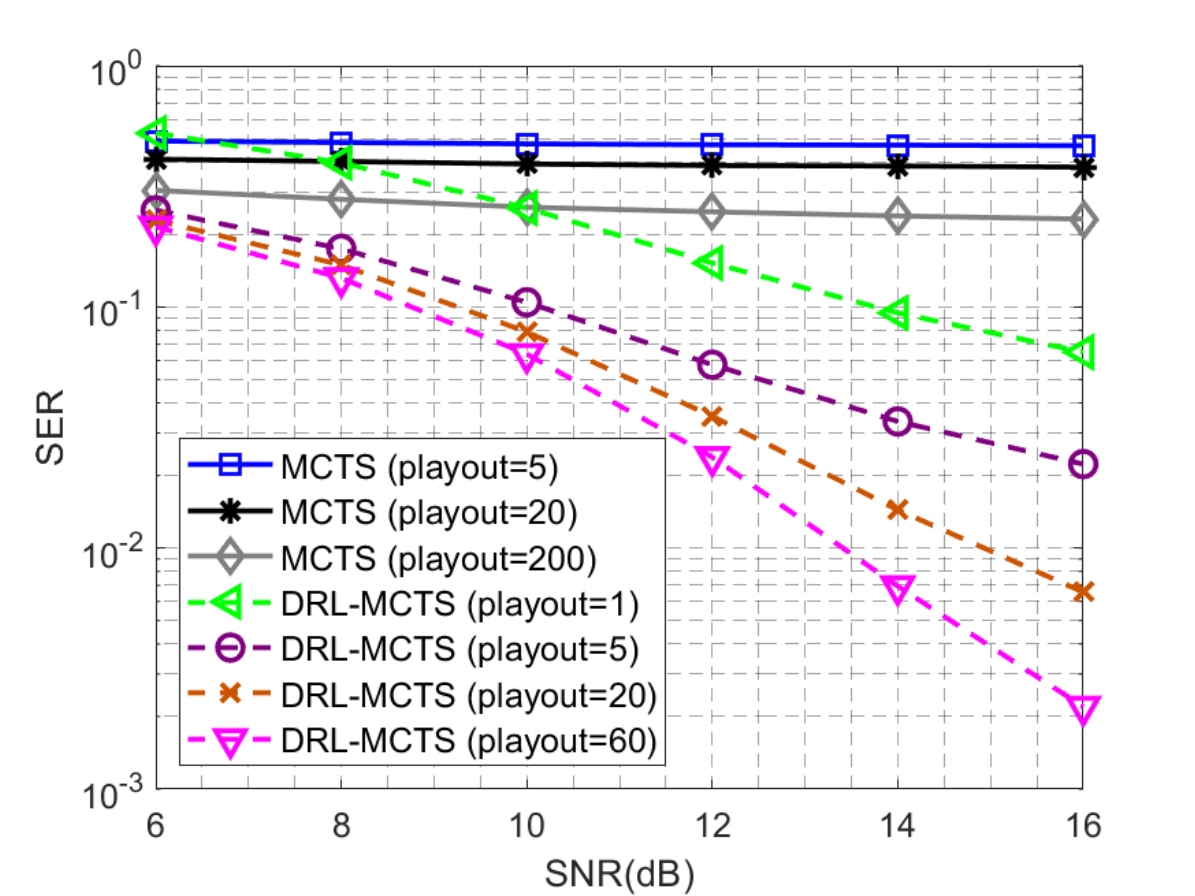}
    \caption{SER performance comparison of MCTS and DRL-MCTS for $8 \times 8$ MIMO with QPSK.}
    \label{fig:playout_8x8QPSK}
\end{figure}

In Fig.~\ref{fig:playout_8x8QPSK}, a similar result is obtained over an $8\times 8$ MIMO channel with QPSK and $\epsilon= 0.08$. We set $c_2 = 1$, $c_{\rm uct}=350$, and $c_{\rm puct}=22$, and playout numbers $5$, $20$, and $200$ for MCTS and $1$, $4$, $20$, and $60$ for DRL-MCTS. Similar trends of curves are exhibited, with lower performance in absolute terms, as compared to the case of BPSK.

\begin{figure}[tb!]
	\centering
	\includegraphics[width=\columnwidth]{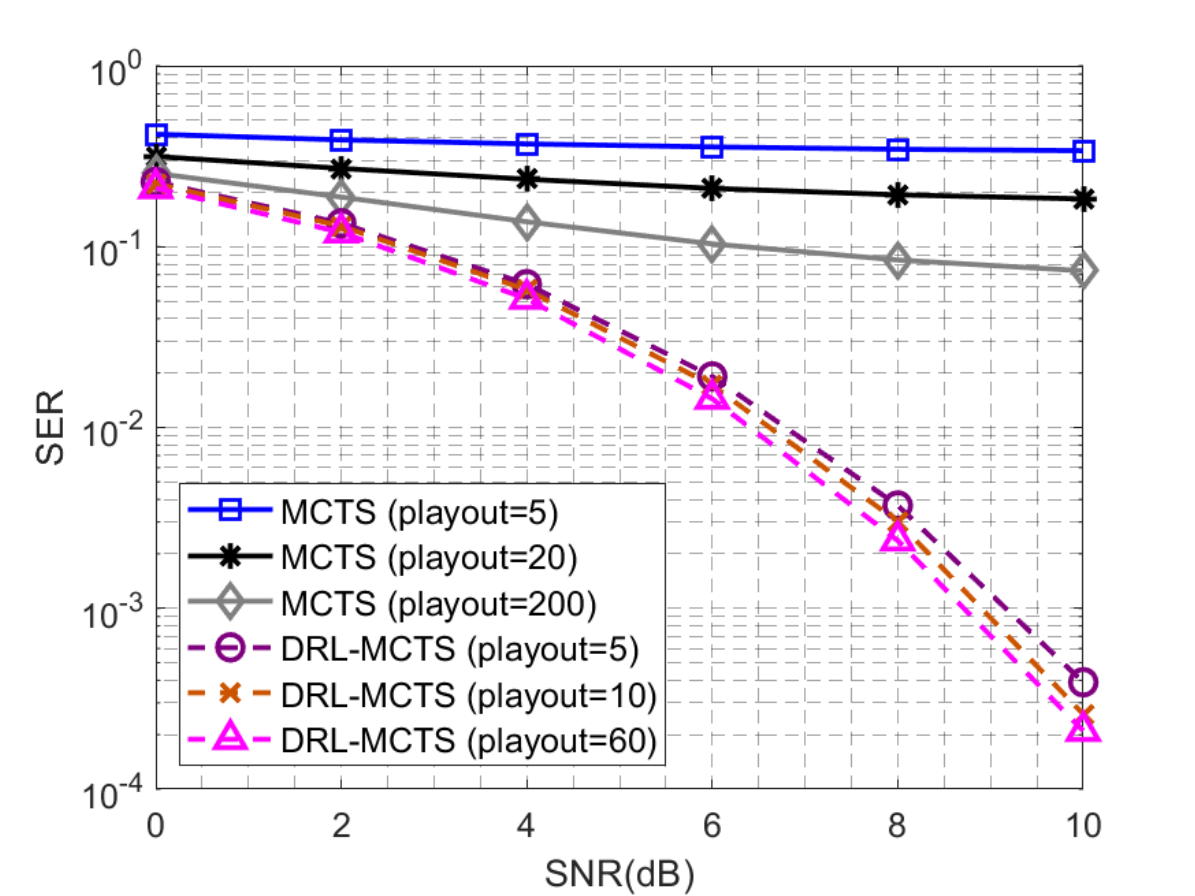}
	\caption{SER performance comparison of MCTS and DRL-MCTS for $8 \times 16$ MIMO with QPSK.}
	\label{fig:playout_8x16QPSK}
\end{figure}

Fig.~\ref{fig:playout_8x16QPSK} examines the performance for an asymmetric $8\times 16$ MIMO channel with QPSK and $\epsilon= 0.08$. We set $c_2 = 1$, $c_{\rm uct}=300$, and $c_{\rm puct}=25$, and playout numbers $5$, $20$, and $200$ for MCTS and $5$, $10$, and $60$ for DRL-MCTS. As can be seen, DRL-MCTS outperforms MCTS at all playout numbers across the entire SNR region. Compared to the previous two scenarios, here, the performance of DRL-MCTS converges faster with fewer numbers of playouts. This may be explained as follows. As the number of antennas increases, the random channel matrix $\mathbf{H}$ converges to a deterministic distribution and the matrix $\mathbf{H}^{\top}\mathbf{H}$ becomes near-diagonal (the ``channel-hardening'' effect) \cite{shaoshi2015mimo}. In addition, when the number of receive antennas is greater than the number of transmit antennas, as in the asymmetric $8\times 16$ channel here, the channel matrix $\mathbf{H}$ tends to be more well-conditioned \cite{shaoshi2015mimo}. This is a favorable condition for traditional detectors such as the sphere decoder (producing extensive tree pruning and therefore higher detection speed and lower complexity) \cite{Ronald2012_3} and MMSE (producing generally satisfactory solutions) \cite{ChangChung2012}, and is similarly so for the tree search-based DRL-MCTS algorithm.

\subsection{DRL-MCTS vs. All Benchmarks Comparison} \label{subsec:results-comparison}

We compare the proposed DRL-MCTS with the following algorithms:
\begin{enumerate}
\item \textbf{MMSE:} The classical linear MMSE detector implemented based on \cite{ChangChung2012}. 
\item \textbf{MCTS:} The detection algorithm described in Sec.~\ref{sec:MCTS}.
\item \textbf{DetNet:} A deep learning based detector proposed in \cite{Samuel2019}.
\item \textbf{DRL:} Detecting MIMO symbols by choosing the action that corresponds to the highest probability from the policy value network output of the DRL agent.
\item \textbf{ML:} The optimal ML detector defined in \eqref{eq:ML_eq}.
\end{enumerate}
In each simulation scenario, we use the same setup as the corresponding scenario described in Sec.~\ref{subsec:results-playout}.

\begin{figure}[tb!]
    \centering
    \includegraphics[width=\columnwidth]{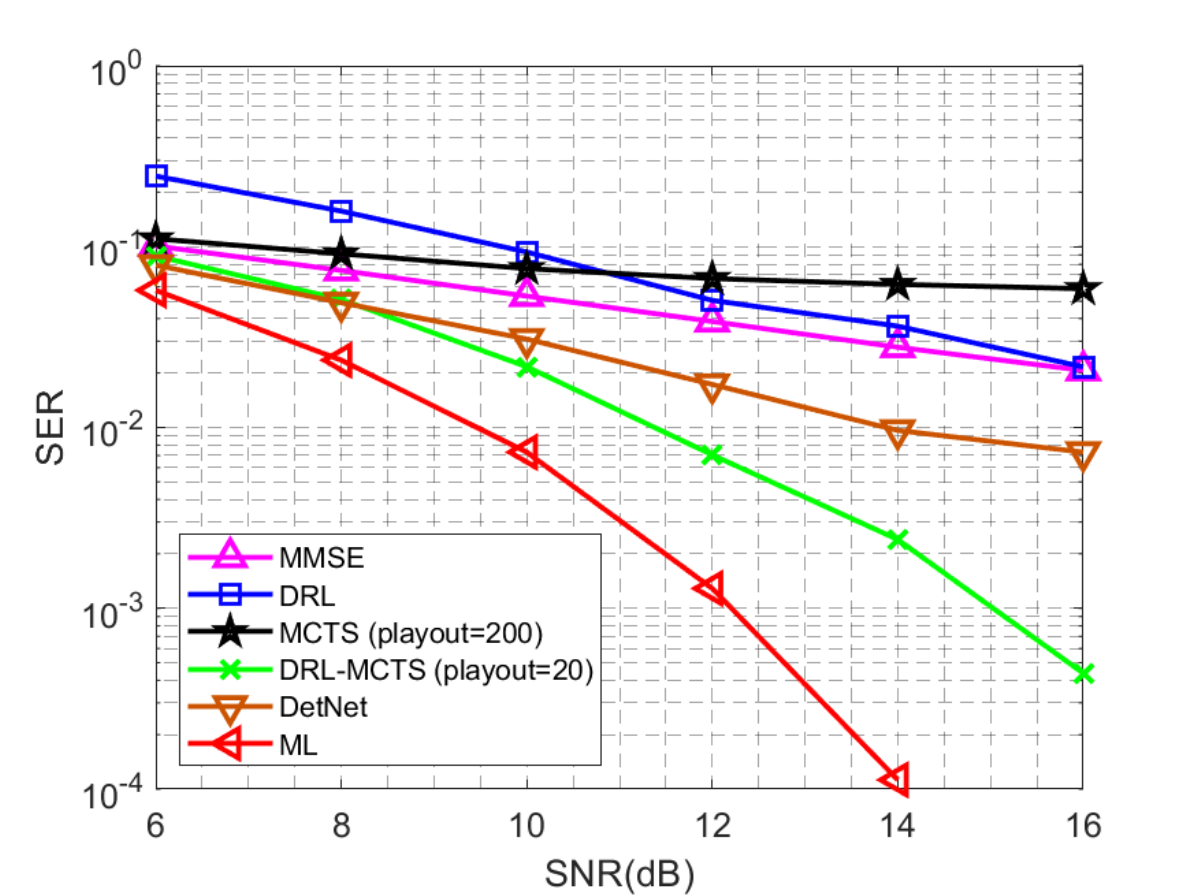}
    \caption{SER performance comparison of different detection algorithms for $8 \times 8$ MIMO with BPSK.}
    \label{fig:8x8BPSK}
\end{figure}

\begin{figure}[tb!]
    \centering
    \includegraphics[width=\columnwidth]{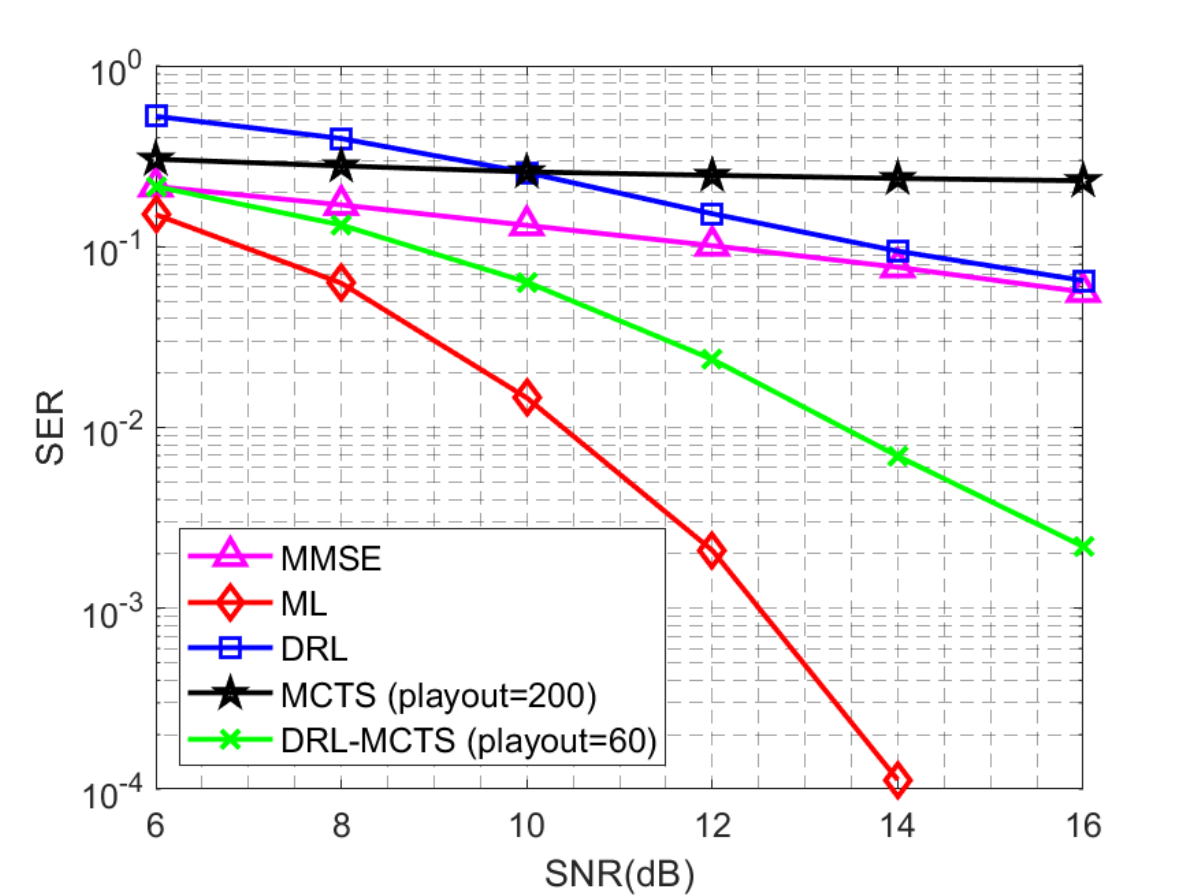}
    \caption{SER performance comparison of different detection algorithms for $8 \times 8$ MIMO with QPSK.}
    \label{fig:8x8QPSK}
\end{figure}

In Fig.~\ref{fig:8x8BPSK}, DRL-MCTS using $20$ playouts is compared against other algorithms over $8\times 8$ BPSK. It can be seen that the proposed DRL-MCTS algorithm significantly outperforms either DRL or MCTS alone. This shows the effectiveness and novelty of combining DRL and MCTS in the proposed method. DRL achieves suffered performance because it directly chooses the action that corresponds to the highest probability from the policy value network output. This, when placed in the context of the DRL-MCTS algorithm, is equivalent to replacing the selection criterion \eqref{eq:puct} by $P^{(i)}$, without utilizing the output of the state value network ($\overline{U}^{(i)}$) and with only one playout ($Z^{(i)} = 1$ and $z^{(i)} = 0$). The omission of the state value network output and the lack of exploration due to one playout lead to the degraded performance of DRL. MCTS also has suffered performance because, as mentioned in Sec.~\ref{sec:MCTS}, it relies on random sampling for simulation, whereas DRL-MCTS relies on learned and guided simulation. More specifically, comparing the selection criteria \eqref{eq:uct} and \eqref{eq:puct} for MCTS and DRL-MCTS, respectively, \eqref{eq:puct} incorporates the state value network output in evaluating the first term and the policy value network output into the second term to enhance tree exploration. DRL-MCTS also outperforms DetNet, with an increasing gain as SNR increases. A similar result is observed in Fig.~\ref{fig:8x8QPSK} for $8\times 8$ QPSK. DetNet was originally proposed for BPSK and thus is not included in the comparison therein.

\begin{figure}[tb!]
	\centering
	\includegraphics[width=\columnwidth]{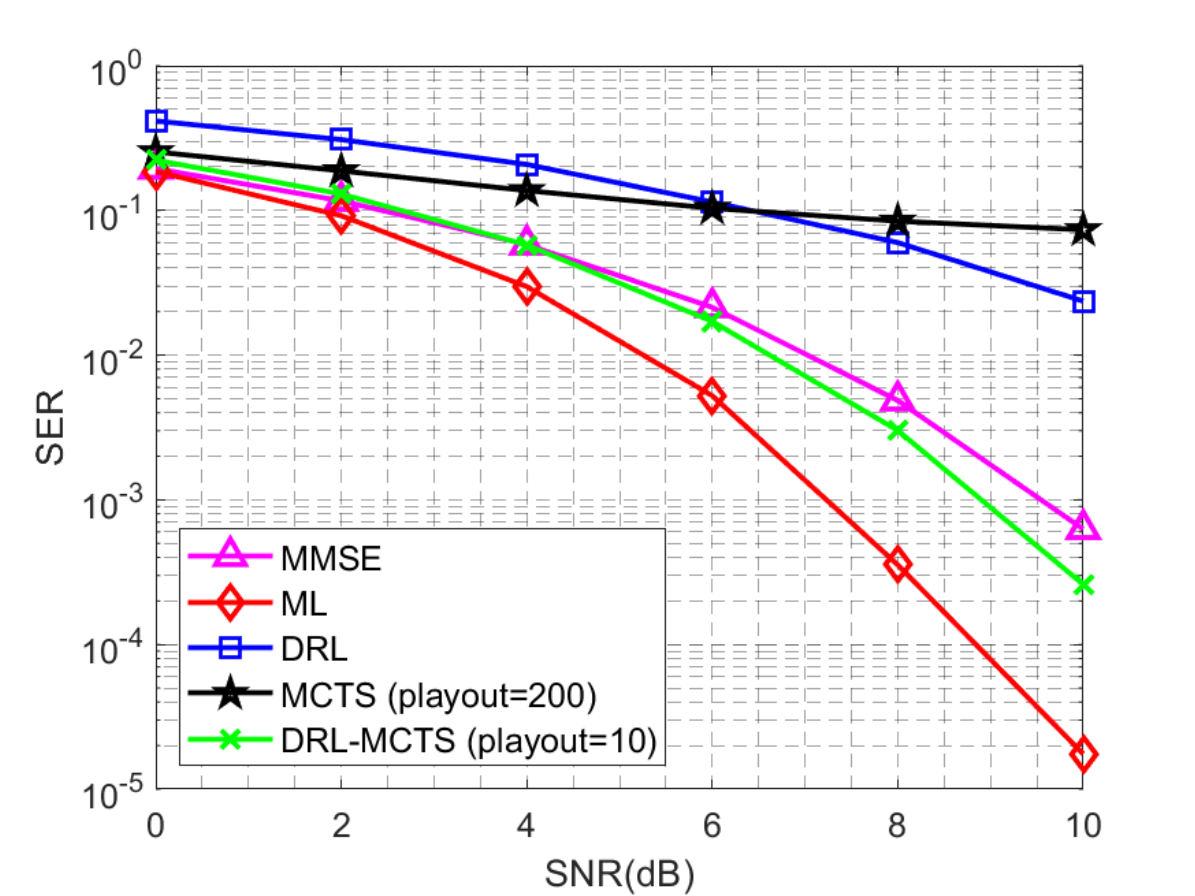}
	\caption{SER performance comparison of different detection algorithms for $8 \times 16$ MIMO with QPSK.}
	\label{fig:8x16QPSK}
\end{figure}

Fig.~\ref{fig:8x16QPSK} shows the result for $8\times 16$ QPSK. The comparative trends of DRL-MCTS, DRL, and MCTS are similar to those in previous scenarios. MMSE achieves satisfactory performance here, since, as discussed, the channel tends to be more well-conditioned in asymmetric channels. Also, as previously mentioned, DRL-MCTS requires only a small number of playouts to achieve good performance in the asymmetric channel.

\subsection{Computational Complexity} \label{subsec:results-complexity}

We evaluate the complexity of all schemes by the runtime complexity. For a fair comparison, all schemes are implemented with Python 3.6 and Tensorflow 1.14, and tested on the same hardware as described in Sec.~\ref{subsec:results-settings}. We simulate sufficient numbers of symbol vectors to get stable results and show the mean detection runtime per symbol vector in the unit of seconds.

\begin{table*}[tb]
\caption{Runtime Comparison of Different Detection Algorithms for $8\times 8$ MIMO with BPSK}
\begin{center}
\begin{tabular}{|c|c|c|c|c|c|c|c|}
\hline
\multicolumn{2}{|c|}{\textbf{Runtime}} & \multicolumn{6}{|c|}{\textbf{SNR}} \\
\cline{3-8} 
\multicolumn{2}{|c|}{\textbf{(seconds)}} & 6 & 8 & 10 & 12 & 14 & 16\\
\hline
\multirow{6}*{\rotatebox{90} {\textbf{Algorithm}}} & ML & $0.0018$ & $0.0023$ & $0.0018$ & $0.0017$ & $0.0022$ & $0.0017$ \\
\cline{2-8}
& MMSE & $4.7$E$-05$ & $4.7$E$-05$ & $4.7$E$-05$ & $4.7$E$-05$ & $4.7$E$-05$ & $4.5$E$-05$ \\
\cline{2-8}
& DRL & $0.0041$ & $0.0034$ & $0.0033$ & $0.0033$ & $0.0035$ & $0.0034$ \\
\cline{2-8}
& MCTS (playout $=200$) & $0.1889$ & $0.1879$ & $0.1873$ & $0.1872$ & $0.1873$ & $0.1872$ \\
\cline{2-8}
& DetNet & $2.2$E$-05$ & $2.1$E$-05$ & $2.0$E$-05$ & $2.0$E$-05$ & $1.9$E$-05$ & $1.9$E$-05$ \\
\cline{2-8}
& DRL-MCTS (playout $=20$) & $0.0457$ & $0.0411$ &  $0.0403$ & $0.0385$ & $0.0365$  & $0.0353$ \\
\hline
\end{tabular}
\label{tab: 8x8BPSK}
\end{center}
\end{table*}

The runtime results for $8\times 8$ BPSK are presented in Table~\ref{tab: 8x8BPSK}. Comparing DRL-MCTS and MCTS, DRL-MCTS not only achieves significantly improved performance as discussed earlier, but also reduced complexity. This can be attributed to the smaller number of playouts required for DRL-MCTS. Comparing DRL-MCTS and DRL, DRL has a lower complexity without the tree search process, but much worse performance. The linear MMSE detector has a low complexity, as expected, at the cost of performance. DetNet is a one-shot DNN-based method and thus runs faster than the proposed algorithm. The search space of ML in this scenario is only $2^8 = 256$, and therefore the complexity of ML is moderate. The runtime for all schemes remain largely unchanged for different SNRs, except for DRL-MCTS where the runtime decreases as SNR increases. This may be explained as follows. When the SNR is higher, DRL-MCTS, similar to traditional tree-search algorithms, generally pursues more focused tree search, and when the same paths are chosen in different playouts, the statistics of previously visited nodes can be reused. This means DRL-MCTS needs to create fewer new child nodes and thus less runtime. In contrast, MCTS randomly chooses and generates new child nodes, and therefore its complexity is unaffected by the SNR factor.

The runtime results for $8\times 8$ QPSK are shown in Table~\ref{tab: 8x8QPSK}. Compared to their counterparts in the BPSK scenario, all algorithms run slower in the QPSK scenario. The comparative trends are similar, even though the complexity of ML here is remarkably higher due to a larger search space that grows quickly with the order of modulation and/or the number of antennas. DRL-MCTS presents a good performance-complexity tradeoff among all schemes.

The runtime results for $8\times 16$ QPSK are shown in Table~\ref{tab: 8x16QPSK}. It is worth noting that, comparing Table~\ref{tab: 8x8QPSK} and Table~\ref{tab: 8x16QPSK}, the complexity of DRL-MCTS significantly reduces while that of DRL and MCTS both increase in this larger antenna system. The increased complexity for DRL and MCTS results from calculating the reward value of larger matrices for each move \cite{jienan2017} and the function operations of higher-dimensional matrices. The main reason for the reduced complexity for DRL-MCTS is the reduced playout number required. This confirms that the number of playouts is the dominating factor of complexity for MCTS-related methods.

\begin{table*}[tb]
\caption{Runtime Comparison of Different Detection Algorithms for $8\times 8$ MIMO with QPSK}
\begin{center}
\begin{tabular}{|c|c|c|c|c|c|c|c|c|}
\hline
\multicolumn{2}{|c|}{\textbf{Runtime}} & \multicolumn{6}{|c|}{\textbf{SNR}} \\
\cline{3-8} 
\multicolumn{2}{|c|}{\textbf{(seconds)}} & 6 & 8 & 10 & 12 & 14 & 16 \\
\hline
\multirow{5}*{\rotatebox{90} {\textbf{Algorithm}}} & ML & $0.4091$ & $0.4873$ & $0.4981$ & $0.5048$ & $0.4950$ & $0.5045$ \\
\cline{2-8}
& MMSE & $5.2$E$-05$ & $5.2$E$-05$ & $5.2$E$-05$ & $5.2$E$-05$ & $5.2$E$-05$ & $5.2$E$-05$ \\
\cline{2-8}
& DRL & $0.0072$ & $0.0067$ & $0.0066$ & $0.0066$ & $0.0066$  & $0.0066$ \\
\cline{2-8}
& MCTS (playout $=200$) & $0.3064$ & $0.3045$ & $0.3037$ & $0.3042$ & $0.3019$ & $0.3018$ \\
\cline{2-8}
& DRL-MCTS (playout $=60$) & $0.2640$ & $0.2546$ &  $0.2455$ & $0.2410$ & $0.2419$ & $0.2401$ \\
\hline
\end{tabular}
\label{tab: 8x8QPSK}
\end{center}
\end{table*}

\begin{table*}[tb]
\caption{Runtime Comparison of Different Detection Algorithms for $8\times 16$ MIMO with QPSK}
\begin{center}
\begin{tabular}{|c|c|c|c|c|c|c|c|c|}
\hline
\multicolumn{2}{|c|}{\textbf{Runtime}} & \multicolumn{6}{|c|}{\textbf{SNR}} \\
\cline{3-8} 
\multicolumn{2}{|c|}{\textbf{(seconds)}} & 0 & 2 & 4 & 6 & 8 & 10 \\
\hline
\multirow{5}*{\rotatebox{90} {\textbf{Algorithm}}} & ML & $0.4734$ & $0.4735$ & $0.4754$ & $0.4779$ & $0.4755$ & $0.4741$ \\
\cline{2-8}
& MMSE & $5.3$E$-05$ & $5.2$E$-05$ & $5.2$E$-05$ & $5.2$E$-05$ & $5.3$E$-05$  & $5.2$E$-05$ \\
\cline{2-8}
& DRL & $0.0075$ & $0.0070$ & $0.0069$ & $0.0070$ & $0.0068$ & $0.0067$ \\
\cline{2-8}
& MCTS (playout $=200$) & $0.3807$ & $0.4039$ & $0.3798$ & $0.3913$ & $0.3919$ &$0.3869$ \\
\cline{2-8}
& DRL-MCTS (playout $=10$) & $0.0735$ & $0.0631$ & $0.0668$ & $0.0594$ & $0.0581$ & $0.0580$ \\
\hline
\end{tabular}
\label{tab: 8x16QPSK}
\end{center}
\end{table*}

\section{Conclusion} \label{sec:conclusion}

In this paper, we proposed a deep reinforcement learning aided Monte Carlo tree search (DRL-MCTS) MIMO detector capable of detecting MIMO symbols under varying channel conditions and different antenna/modulation configurations. By using DRL, we avoid the labeling dilemma encountered in other learning-based detectors using supervised learning. The proposed scheme innovatively combines DRL and MCTS to enhance MCTS with a learned and guided tree search process enabled by DRL. Simulation results demonstrated that incorporating DRL into the MCTS framework significantly increases the detection accuracy and search efficiency of the original MCTS detector. The proposed DRL-MCTS detector was also shown to outperform traditional linear detector and DNN-based detector. Extensive and insightful discussions on the performance and complexity of various schemes were provided.

\bibliographystyle{IEEEtran}
\bibliography{IEEEabrv,references}

\begin{thebibliography}{10}
\providecommand{\url}[1]{#1}
\csname url@samestyle\endcsname
\providecommand{\newblock}{\relax}
\providecommand{\bibinfo}[2]{#2}
\providecommand{\BIBentrySTDinterwordspacing}{\spaceskip=0pt\relax}
\providecommand{\BIBentryALTinterwordstretchfactor}{4}
\providecommand{\BIBentryALTinterwordspacing}{\spaceskip=\fontdimen2\font plus
\BIBentryALTinterwordstretchfactor\fontdimen3\font minus
  \fontdimen4\font\relax}
\providecommand{\BIBforeignlanguage}[2]{{%
\expandafter\ifx\csname l@#1\endcsname\relax
\typeout{** WARNING: IEEEtran.bst: No hyphenation pattern has been}%
\typeout{** loaded for the language `#1'. Using the pattern for}%
\typeout{** the default language instead.}%
\else
\language=\csname l@#1\endcsname
\fi
#2}}
\providecommand{\BIBdecl}{\relax}
\BIBdecl

\bibitem{verdu1989}
S.~Verdu, ``Computational complexity of optimum multiuser detection,''
  \emph{Algorithmica}, vol.~4, no.~1, pp. 303--312, Dec. 1989.

\bibitem{shaoshi2015mimo}
S.~Yang and L.~Hanzo, ``Fifty years of {MIMO} detection: The road to
  large-scale {MIMOs},'' \emph{IEEE Commun. Surveys Tuts.}, vol.~17, no.~4, pp.
  1941--1988, Fourth Quarter 2015.

\bibitem{cheng2007zf}
C.~Wang, E.~K. Au, R.~D. Murch, W.~H. Mow, R.~S. Cheng, and V.~Lau, ``On the
  performance of the {MIMO} zero-forcing receiver in the presence of channel
  estimation error,'' \emph{IEEE Trans. Wireless Commun.}, vol.~6, no.~3, pp.
  805--810, Mar. 2007.

\bibitem{ChangChung2012}
R.~Y. Chang and W.-H. Chung, ``Low-complexity {MIMO} detection based on
  post-equalization subspace search,'' \emph{IEEE Trans. Veh. Technol.},
  vol.~61, no.~1, pp. 375--380, Jan. 2012.

\bibitem{marcenko1967channelharden}
V.~A. Marčenko and L.~A. Pastur, ``Distribution of eigenvalues for some sets
  of random matrices,'' \emph{Math USSR Shornik}, vol.~1, no.~4, pp. 457--483,
  Nov. 1967.

\bibitem{Michael2013}
M.~Wu, B.~Yin, A.~Vosoughi, C.~Studer, J.~R. Cavallaro, and C.~Dick,
  ``Approximate matrix inversion for high-throughput data detection in the
  large-scale {MIMO} uplink,'' \emph{in 2013 IEEE International Symposium on
  Circuits and Systems (ISCAS)}, pp. 2155--2158, May. 2013.

\bibitem{dengkui2015}
D.~Zhu, B.~Li, and P.~Liang, ``On the matrix inversion approximation based on
  {Neumann} series in massive {MIMO} systems,'' \emph{in 2015 IEEE
  International Conference on Communications (ICC)}, pp. 1763--1769, Jun. 2015.

\bibitem{byunggi2015}
B.~Kang, J.-H. Yoon, and J.~Park, ``Low complexity massive {MIMO} detection
  architecture based on {Neumann} method,'' \emph{2015 International SoC Design
  Conference (ISOCC)}, pp. 293--294, Nov. 2015.

\bibitem{andrew2005}
J.~G. {Andrews}, ``Interference cancellation for cellular systems: a
  contemporary overview,'' \emph{IEEE Wireless Commun.}, vol.~12, no.~2, pp.
  19--29, Apr. 2005.

\bibitem{Bittner2006}
S.~Bittner, E.~Zimmermann, and G.~Fettweis, ``Low complexity soft interference
  cancellation for {MIMO}-systems,'' \emph{in 2006 IEEE 63rd Vehicular
  Technology Conference}, pp. 1993--1997, May 2006.

\bibitem{Jinho2009}
J.~Choi and H.~Nguyen, ``{SIC-Based} detection with list and lattice reduction
  for {MIMO} channels,'' \emph{IEEE Trans. Veh. Technol.}, vol.~58, no.~7, pp.
  3786--3790, Sep. 2009.

\bibitem{Peng2010}
P.~Li and R.~D. Murch, ``Multiple output selection-{LAS} algorithm in large
  {MIMO} systems,'' \emph{IEEE Commun. Lett.}, vol.~14, no.~5, pp. 399--401,
  May 2010.

\bibitem{Tanumay2010}
T.~Datta, N.~Srinidhi, A.~Chockalingam, and B.~S. Rajan, ``Random-restart
  reactive tabu search algorithm for detection in large-{MIMO} systems,''
  \emph{IEEE Commun. Lett.}, vol.~14, no.~12, pp. 1107--1109, Oct. 2010.

\bibitem{Jong2005}
Y.~de~Jong and T.~Willink, ``Iterative tree search detection for {MIMO}
  wireless systems,'' \emph{IEEE Trans. Commun.}, vol.~53, no.~6, pp. 930--935,
  Jun. 2005.

\bibitem{Atsushi2008}
A.~Okawado, R.~Matsumoto, and T.~Uyematsu, ``Near {ML} detection using
  {Dijkstra's} algorithm with bounded list size over {MIMO} channels,''
  \emph{in 2008 IEEE International Symposium on Inform. Theory}, pp.
  2022--2025, Jul. 2008.

\bibitem{Ronald2012}
R.~Y. Chang and W.-H. Chung, ``Best-first tree search with probabilistic node
  ordering for {MIMO} detection: Generalization and performance-complexity
  tradeoff,'' \emph{IEEE Trans. Wireless Commun.}, vol.~11, no.~2, pp.
  780--789, Feb. 2012.

\bibitem{Ronald2012_2}
R.~Y. Chang, W.-H. Chung, and S.-J. Lin, ``A* algorithm inspired
  memory-efficient detection for {MIMO} systems,'' \emph{IEEE Wireless Commun.
  Lett.}, vol.~1, no.~5, pp. 508--511, Jul. 2012.

\bibitem{Ronald2012_3}
R.~Y. Chang, S.-J. Lin, and W.-H. Chung, ``Efficient implementation of the
  {MIMO} sphere detector: Architecture and complexity analysis,'' \emph{IEEE
  Trans. Veh. Technol}, vol.~61, no.~7, pp. 3289--3294, Jun. 2012.

\bibitem{jienan2017}
J.~{Chen}, C.~{Fei}, H.~{Lu}, G.~E. {Sobelman}, and J.~{Hu}, ``Hardware
  efficient massive {MIMO} detector based on the {Monte Carlo} tree search
  method,'' \emph{IEEE Journal on Emerging and Selected Topics in Circuits and
  Systems}, vol.~7, no.~4, pp. 523--533, Aug. 2017.

\bibitem{5503188}
P.~{Som}, T.~{Datta}, A.~{Chockalingam}, and B.~S. {Rajan}, ``Improved
  {large-MIMO} detection based on damped belief propagation,'' in \emph{2010
  IEEE Information Theory Workshop on Information Theory (ITW 2010, Cairo)},
  Jan. 2010, pp. 1--5.

\bibitem{Junmei2015}
J.~Yang, C.~Zhang, X.~Liang, S.~Xu, and X.~You, ``Improved symbol-based belief
  propagation detection for large-scale {MIMO},'' \emph{in 2015 IEEE Workshop
  on Signal Processing Systems (SiPS)}, Oct. 2015.

\bibitem{Jalden2008}
J.~{Jalden} and B.~{Ottersten}, ``The diversity order of the semidefinite
  relaxation detector,'' \emph{IEEE Trans. Inf. Theory}, vol.~54, no.~4, pp.
  1406--1422, Mar. 2008.

\bibitem{Jeon2015}
C.~{Jeon}, R.~{Ghods}, A.~{Maleki}, and C.~{Studer}, ``Optimality of large
  {MIMO} detection via approximate message passing,'' in \emph{2015 IEEE
  International Symposium on Information Theory (ISIT)}, Jun. 2015, pp.
  1227--1231.

\bibitem{Chen2019}
Q.~{Chen}, S.~{Zhang}, S.~{Xu}, and S.~{Cao}, ``Efficient {MIMO} detection with
  imperfect channel knowledge - a deep learning approach,'' in \emph{2019 IEEE
  Wireless Communications and Networking Conference (WCNC)}, Apr. 2019, pp.
  1--6.

\bibitem{Samuel2019}
N.~Samuel, T.~Diskin, and A.~Wiesel, ``Learning to detect,'' \emph{IEEE Trans.
  Signal Process.}, vol.~67, no.~10, pp. 2554--2564, May 2019.

\bibitem{Vincent2018}
V.~{Corlay}, J.~J. {Boutros}, P.~{Ciblat}, and L.~{Brunel}, ``Multilevel {MIMO}
  detection with deep learning,'' in \emph{2018 52nd Asilomar Conference on
  Signals, Systems, and Computers}, Oct. 2018, pp. 1805--1809.

\bibitem{mohammad2020}
A.~{Mohammad}, C.~{Masouros}, and Y.~{Andreopoulos}, ``Complexity-scalable
  neural-network-based {MIMO} detection with learnable weight scaling,''
  \emph{IEEE Trans. Commun}, vol.~68, no.~10, pp. 6101--6113, 2020.

\bibitem{jin2020}
X.~{Jin} and H.~{Kim}, ``Parallel deep learning detection network in the {MIMO}
  channel,'' \emph{IEEE Commun. Lett.}, vol.~24, no.~1, pp. 126--130, 2020.

\bibitem{Nir2020}
N.~Shlezinger, R.~Fu, and Y.~C. Eldar, ``{DeepSIC}: Deep soft interference
  cancellation for multiuser {MIMO} detection,'' \emph{IEEE Trans. Wireless
  Commun. (Early Access)}, 2020.

\bibitem{Jianyong2020}
J.~Sun, Y.~Zhang, J.~Xue, and Z.~Xu, ``Learning to search for {MIMO}
  detection,'' \emph{IEEE Trans. Wireless Commun.}, vol.~19, no.~11, pp.
  7571--7584, Nov. 2020.

\bibitem{Tan2020}
X.~{Tan}, W.~{Xu}, K.~{Sun}, Y.~{Xu}, Y.~{Be'ery}, X.~{You}, and C.~{Zhang},
  ``Improving massive {MIMO} message passing detectors with deep neural
  network,'' \emph{IEEE Trans. Veh. Technol.}, vol.~69, no.~2, pp. 1267--1280,
  Dec. 2020.

\bibitem{Hengtao2018}
H.~{He}, C.~{Wen}, S.~{Jin}, and G.~Y. {Li}, ``A model-driven deep learning
  network for {MIMO} detection,'' in \emph{2018 IEEE Global Conference on
  Signal and Information Processing (GlobalSIP)}, 2018, pp. 584--588.

\bibitem{cameron2012mcts}
C.~B. Browne, E.~Powle, D.~Whitehouse, S.~M. Lucas, P.~I. Cowling,
  P.~Rohlfshagen, S.~Tavener, D.~Perez, S.~Samothrakis, and S.~Colton, ``A
  survey of {Monte Carlo} tree search methods,'' \emph{{IEEE} Trans. Comput.
  Intell. {AI} in Games}, vol.~4, no.~1, pp. 1--43, Mar. 2012.

\bibitem{David2017alphago}
D.~Silver, J.~Schrittwieser, K.~Simonyan, I.~Antonoglou, A.~Huang, A.~Guez,
  T.~Hubert, L.~Baker, M.~Lai, A.~Bolton, Y.~Chen, T.~Lillicrap, F.~Hui,
  L.~Sifre, G.~van~den Driessche, T.~Graepel, and D.~Hassabis, ``Mastering the
  game of {Go} without human knowledge,'' \emph{Nature}, vol. 550, pp.
  354--359, Oct. 2017.

\bibitem{konda2000}
V.~Konda and J.~Tsitsiklis, ``Actor-critic algorithms,'' in \emph{SIAM Journal
  on Control and Optimization}.\hskip 1em plus 0.5em minus 0.4em\relax MIT
  Press, 2000, pp. 1008--1014.

\bibitem{volodymyr2016}
\BIBentryALTinterwordspacing
V.~Mnih, A.~P. Badia, M.~Mirza, A.~Graves, T.~P. Lillicrap, T.~Harley,
  D.~Silver, and K.~Kavukcuoglu, ``Asynchronous methods for deep reinforcement
  learning,'' \emph{CoRR}, vol. abs/1602.01783, 2016. [Online]. Available:
  \url{http://arxiv.org/abs/1602.01783}
\BIBentrySTDinterwordspacing

\bibitem{JeonLee2020}
Y.-S. Jeon, N.~Lee, and H.~V. Poor, ``Robust data detection for {MIMO} systems
  with one-bit {ADCs}: A reinforcement learning approach,'' \emph{IEEE Trans.
  Wireless Commun.}, vol.~19, no.~3, pp. 1663--1676, Mar. 2020.

\end{thebibliography}

\end{document}